\documentclass[onecolumn,nobibnotes]{revtex4}

\usepackage{amsmath}
\usepackage{graphicx}
\usepackage{float}
\usepackage{wrapfig}
\usepackage{multirow}
\usepackage{tabularx}
\usepackage{array}
\usepackage{url}
\usepackage{tikz}
\usepackage{comment}
\usepackage{ulem}
\usetikzlibrary{positioning}
\usepackage{graphicx,color}
\usepackage{hyperref}
\hypersetup{
    colorlinks=true,
    linkcolor=black,
    citecolor=black,
    }
\def\lsim{\raise0.3ex\hbox{$<$\kern-0.75em\raise-1.1ex\hbox{$\sim$}}}
\def\gsim{\raise0.3ex\hbox{$>$\kern-0.75em\raise-1.1ex\hbox{$\sim$}}}

\def\beq{\begin{equation}}
\def\eeq{\end{equation}}
\def\beqa{\begin{eqnarray}}
\def\eeqa{\end{eqnarray}}

\def\gappeq{\mathrel{\rlap {\raise.5ex\hbox{$>$}}
{\lower.5ex\hbox{$\sim$}}}}

\def\lappeq{\mathrel{\rlap{\raise.5ex\hbox{$<$}}
{\lower.5ex\hbox{$\sim$}}}}

\def\Toprel#1\over#2{\mathrel{\mathop{#2}\limits^{#1}}}

\begin{document}

\title{KNO scaling violation: the role of the fragmentation region}

\author{R. Terra$^1$, A. V. Giannini$^{2,3}$ and F. S. Navarra$^1$ }
\affiliation{$^1$Instituto de F\'{\i}sica, Universidade de S\~{a}o Paulo, 
Rua do Mat\~ao, 1371, CEP 05508-090,  S\~{a}o Paulo, SP, Brazil\\
$^2$ Faculdade de Ciências Exatas e Tecnologia, Universidade Federal da 
Grande Dourados, CEP 79804-970,  Dourados, MS, Brazil\\
$^3$ Departamento de F\'isica, Universidade do Estado de Santa Catarina, 
89219-710 Joinville, SC, Brazil
}

\begin{abstract}

This work is motivated by two independent considerations. The first is
that KNO scaling is observed in the central rapidity region ($|\eta|<0.5$, 
for example) and violated when the fragmentation region is included 
($|\eta|<3.4$, for example). The
second is that, in the CGC framework, the $k_T$-factorization    
formalism is more appropriate to describe particle production in the central 
region and the hybrid formalism
is more appropriate for the fragmentation region. The former contains small-x 
gluon interactions, where the latter introduce also collinear gluons and valence
quarks in the dynamics. These two considerations put together raise the natural 
question: are collinear gluons and valence quarks responsible for KNO scaling  
violation? To answer this question we build a two-component CGC model which is 
able to reproduce the measured pseudorapidity distributions over all 
the available range as well as multiplicity distributions  in pp collisions at 
LHC energies. With this model we find that KNO scaling occurs when the interaction 
is dominated by small-x gluons but is broken by the interactions including 
small-x gluons, collinear gluons and valence quarks.
\end{abstract}

\maketitle

\section{Introduction}

\subsection{KNO scaling and its violation}

Multiplicity distributions of produced particles in high-energy hadronic
collisions have played a fundamental role in understanding the dynamics of
soft particle production and the transition between non-perturbative and
perturbative regimes of Quantum Chromodynamics (QCD). For reviews, see
\cite{mult1,mult2}. One of the most remarkable 
observations in this field was the discovery of an universal
scaling behavior of charged-particle multiplicity distributions, known as
Koba--Nielsen--Olesen (KNO) scaling, proposed in 1972 \cite{KNO}. The central 
idea of KNO scaling is 
that the probability distribution for producing $n$ particles in a collision
does not depend explicitly on the collision energy $\sqrt{s}$ when expressed
in terms of the multiplicity normalized by its average value. In other words,
the shape of the multiplicity distribution becomes energy independent after
rescaling the horizontal and vertical axes. Defining the scaling variable 
$z=n/\langle n \rangle$, KNO scaling states that
\begin{equation}
\Psi(z)=\langle n\rangle(s) \,  P_n(s)
\end{equation}
is a universal function independent of the center-of-mass energy, i.e., 
$\Psi(z,s)=\Psi(z)$. 

This property suggested that the complicated dynamics of particle
production might be governed by a simple self-similar mechanism~\cite{Braun:1998ty}.
The average multiplicity $\langle n \rangle$ increases with collision energy,
but KNO scaling implies that fluctuations around the mean increase in a
precisely correlated way. The probability of producing a multiplicity $n$
at energy $\sqrt{s}$ depends only on the ratio $n/\langle n\rangle$.
Equivalently, all normalized moments of the multiplicity distribution,
\begin{equation}
C_q=\frac{\langle n^q\rangle}{\langle n\rangle^q},
\end{equation}
should be independent of the energy. Experimentally, this means that quantities
such as the dispersion-to-mean ratio,
${D^2}/{\langle n\rangle^2} =[\langle n^2\rangle-\langle n\rangle^2]
/{\langle n\rangle^2}$
should remain constant as the collision energy increases.
Early measurements at the Intersecting Storage Rings (ISR) at CERN showed
approximately energy-independent multiplicity distributions when presented
in the KNO form. These observations provided strong support for the idea
that soft hadronic interactions possess a universal scaling behavior
\cite{mult1,mult2}.

However, as higher energies became available at the SPS, Tevatron and,
especially, at the Large Hadron Collider (LHC), deviations from KNO scaling
were observed \cite{cms11,ALICE_17,atlas22,atlas23}. 
The increasing violation of this scaling law revealed new
features of particle production dynamics.
The first indications of KNO violation appeared when multiplicity
distributions were measured at energies significantly above the ISR range.
Experiments such as UA5 \cite{UA5} at the CERN SPS observed that the    
high-multiplicity tail of the distribution became progressively broader 
than predicted by KNO scaling.

At the Tevatron and the LHC, the violation became evident. Measurements by
CDF, ATLAS, CMS and ALICE demonstrated that the normalized multiplicity
distribution evolves with energy, particularly at large values of $z$.
The increase of the high-multiplicity tail indicates that rare events with
particle production much larger than the average become increasingly
important at high energies. The violation is also visible in the normalized 
moments. Instead of being constant, quantities such as $C_q$
increase with energy, showing that multiplicity fluctuations become stronger
as the collision energy grows. For a recent review on the energy dependence 
of the moments $C_q$, see Ref. \cite{germano2}.

As it will be discussed below, the modern interpretation is that KNO 
violation is a consequence of several interrelated effects: perturbative 
QCD evolution, multiple parton interactions,  multi-Pomeron exchange, and 
fluctuations of the gluonic structure of hadrons. In this work we add to 
this list one more source of KNO violation: production from the fragmentation
region.

\subsection{Theoretical explanations of KNO violation}

Several mechanisms, which will be briefly presented next, have been proposed to explain the breakdown of KNO
scaling. They are not mutually exclusive, and modern interpretations usually
combine several of them.

\subsubsection{QCD parton cascades and running coupling effects}

The original KNO picture was formulated before the development of modern
QCD. In perturbative QCD, particle production is described by a cascade of
parton branchings, $q\rightarrow qg$, $ g\rightarrow gg$ and 
$g\rightarrow q\bar q. $
Because the strong coupling constant decreases with increasing momentum scale,
$
\alpha_s(Q^2)={1}/[b_0\ln(Q^2/\Lambda_{\rm QCD}^2)],
$
the evolution of parton showers introduces an energy dependence into
multiplicity fluctuations.

The Double Logarithmic Approximation (DLA) and later Modified Leading
Logarithmic Approximation (MLLA) calculations predicted violations of KNO
scaling due to the QCD evolution of parton cascades
\cite{Mueller,Dremin,Dokshitzer}. The main prediction is that higher-order multiplicity moments slowly evolve
with energy, in qualitative agreement with observations.

\subsubsection{Negative Binomial distributions and clan models}

A phenomenologically successful description of multiplicity distributions is
provided by the Negative Binomial Distribution (NBD):
\begin{equation}
P_n =
\frac{\Gamma(n+k)}
{\Gamma(n+1)\Gamma(k)}
\left(\frac{\langle n\rangle}{\langle n\rangle+k}\right)^n
\left(\frac{k}{\langle n\rangle+k}\right)^k ,
\end{equation}
where the parameter $k$ controls the strength of multiplicity fluctuations.
In the KNO limit, the NBD approaches a scaling function. However, the
energy dependence of $k$ introduces a violation of exact KNO scaling.

The NBD interpretation was developed within the framework of clan models,
where particle production proceeds through two stages: the formation of
clusters (or ``clans'') followed by their decay into final-state particles.
The increase of the number of clans and their correlations with increasing
energy naturally produces a broader multiplicity distribution
\cite{gio86}.

\subsubsection{Multiple parton interactions}

At collider energies, especially at the LHC, a single proton--proton
collision may contain several semi-hard parton--parton interactions.
Multiple Parton Interactions (MPI) increase the probability of producing
events with unusually high multiplicities. Monte Carlo event generators such as
\textsc{PYTHIA} and \textsc{Herwig} describe KNO violation largely through
the increasing importance of MPI, together with color reconnection effects.
The growth of the high-$z$ tail of the multiplicity distribution is a
natural consequence of these additional particle-production mechanisms 
\cite{mpi}. 

\subsubsection{Multi-Pomeron exchange models}

Another important class of explanations is based on Regge theory and
Pomeron exchange. In this approach, soft hadronic interactions are described
as exchanges of multiple Pomerons, each producing a particle cascade.
At low energies a single Pomeron contribution may approximately satisfy KNO
scaling. At higher energies, however, the probability of multiple Pomeron
exchange increases, producing additional fluctuations and therefore violating
KNO scaling. These ideas were developed in the Dual Parton Model and related 
approaches \cite{DPM}.

\subsubsection{Color Glass Condensate and saturation approaches}

At very high energies, the small-$x$ gluon density of hadrons becomes very
large, and nonlinear QCD effects become important. The Color Glass        
Condensate (CGC) framework describes this regime through gluon saturation 
\cite{thecgc,saturation_physics}. In this picture, multiplicity fluctuations 
are controlled by fluctuations
of the saturation scale, $Q_s^2(x)$, which evolves with energy. The 
increasing importance of gluon density fluctuations naturally leads to 
deviations from KNO scaling.

In this  framework, KNO scaling and its      
violation are understood as consequences of the dynamics of small-x gluons, 
saturation, and event-by-event fluctuations of the color charge density
\cite{dumi}.
Approximate KNO scaling arises because particle production is 
dominated by many independent Glasma flux tubes whose multiplicity       
distribution is a negative binomial, with both the mean multiplicity and 
the fluctuation parameter controlled by the saturation scale $Q_s$. 
Exact KNO scaling requires the effective action for small-x color sources 
to remain nearly Gaussian under quantum evolution~\cite{dumi}. 
Non-Gaussian corrections generated by JIMWLK evolution, fluctuations of the 
saturation momentum, higher-order correlations among color sources, 
and rare-event dynamics would broaden the high-multiplicity tail and 
lead to scaling violations.

\subsection{The role of the fragmentation region} 

Since long time ago \cite{vanhove,fowler}, 
particle production in high-energy hadronic collisions 
(such as pp, pA, or AA collisions), is commonly divided into two broad 
kinematic regions: the fragmentation region (large rapidities) and the 
central region (smaller rapidities).

The fragmentation region is the region close to the beam rapidity. 
Particles in this region originate mainly from 
the breakup (fragmentation) of the incoming hadrons, they retain a significant 
fraction of the beam momentum and are strongly correlated with the quantum  
numbers of the incoming hadrons. Here, particle production is dominated by 
valence quarks, beam remnants, string fragmentation near the projectile and 
diquark fragmentation. Particles ``remember'' the 
incoming hadrons. 
For example, the baryon number tends to remain near beam rapidity and leading particle 
effects are important.

The central region is defined to be around $y \simeq 0$ in the center-of-mass 
frame. Particles here are produced far from either beam rapidity, and by 
kinematical constraints, carry only a small fraction of the initial 
longitudinal momentum. The production mechanism is dominated by gluon-gluon interactions, 
sea quarks, multiple parton interactions, color field fragmentation, minijets 
and jets. The observed particles originate from newly produced partons 
rather than directly inherited from the projectiles.

A striking feature revealed by the LHC measurements is that KNO scaling is 
approximately valid in the central region and it is violated in the    
fragmentation region \cite{ALICE_pp_eta_08,cms11,ALICE_17,ALICE_wide}. How does this experimental finding fit in the above listed explanations of KNO violation?

The Color Glass Condensate predicts that gluon production is correlated only  
over a finite interval in rapidity. As long as the rapidity 
window is smaller than this correlation length, the emitted particles originate from 
essentially the same ensemble of color fields, and KNO scaling is       
approximately preserved. Once the rapidity interval becomes much larger 
than the correlation length, independent fluctuations from different 
rapidity slices are added together, and the KNO scaling function changes.
This property links the onset of KNO scaling violations directly to 
the finite rapidity correlation length of the Glasma and the rapidity 
evolution of the CGC effective theory \cite{dumi}. As a consequence, particles
produced in a narrow rapidity slice will have different properties from those 
of the particles produced in a broad rapidity interval. This could explain the 
experimental data. 
The CGC picture provides a more dynamical explanation than older phenomenological 
models, in which the violation was attributed mainly to the coexistence of 
soft and semihard particle-production mechanisms \cite{henrique}.

The ideas sketched above give us a qualitative understanding of the 
features of particle production in the CGC framework. However, they were not
yet converted into a calculation algorithm able to reproduce the data in 
full detail. In this work we revisit this question and address the data on
KNO scaling using the CGC formalism in the central region and in the forward 
region.

\section{Formalism}

The $k_T$-factorization and hybrid formalisms are the two most widely used 
approaches for calculating particle production within the CGC \cite{alba14,khar05,rcbk,hibr,pA_karzeev,dunga,nara_26,yuri_D}. Although both are derived from the same      
underlying  effective theory, they rely on different kinematic approximations 
and therefore have different domains of validity. 

The $k_T$-factorization formalism \cite{kov02} assumes that both projectiles  
are in the small-x regime, i.e., typically $x_1\ll 1$ and $x_2 \ll 1$. Each 
hadron is then described by an unintegrated gluon distribution (UGD) 
\cite{khar05}.  At forward rapidities the
kinematic conditions above are not satisfied. For $y \approx 5$,  
$x_1 = [p_T/\sqrt{s}] \, \exp(y)$ can easily become $0.1$ or larger. 
At such  $x$, the projectile is no longer gluon dominated and the formalism 
breaks down. 

The hybrid formalism was developed later to describe forward particle 
production \cite{alba14}.  Here, typically $x_1 \simeq 0.1$ and 
$x_2 \ll 1$. The projectile is thus dilute and the target is in a denser 
state .
So, the non-linear dynamics associated with the CGC effective field theory 
is encoded in the target, while the projectile state is described by  
integrated parton distributions that are solution of linear evolution   
equations such as DGLAP. These partons scatter eikonally from the dense 
gluon field of 
the target. At LHC energies, when $y \approx 0$, we have typically           
$x_1 \simeq x_2 \simeq 10^{-4}$. In this situation both hadrons are expected 
to be found in a dense state and the hybrid approximation (which assumes one 
dilute projectile) 
is no longer justified. The same analysis is valid for the backward particle 
production by considering that the target is probed at large-$x$ and the 
projectile at low-$x$.

As we can see, it is assumed a separation between the validity domains of 
both formalisms. Usually, in practical calculations, we are interested only  
in one domain or in the other. However, a description of the experimentally 
measured rapidity distribution in different regions of the phase space is 
highly desirable. In order to do so, we consider a two-component CGC model 
composed by the $k_T$-factorization and the hybrid formalism, in the same 
spirit of the Glauber model~\cite{Miller:2007ri}. 
Since the region of validity of each component is defined 
	in terms of asymptotic regions, we employ a (pseudo)rapidity-dependent 
	weight function to smoothly interpolate both formalisms over the phase 
	space, as described below.

\subsection{$k_T$-factorization}
 
In order to compute the rapidity and transverse momentum distributions of the  
particles produced in the final state near midrapidity, we use the 
$k_T$-factorization, \cite{alba14,khar05,rcbk}, where the single-inclusive (small-x) gluon production is given by:
\begin{eqnarray}
\frac{d N^{A+B \rightarrow g}}{dy \, d^2p_T \, d^2 \mathbf{r_{\perp}}} = \frac{K_{kT}}{\sigma_{s}}
 \, \frac{2}{C_F\text{ } p_T^2}
\int^{p_T} \frac{d^2k_T}{4} \int d^2b \text{ } \alpha_s(Q) \, 
\phi_p \bigg(\frac{|p_T+k_T|}{2},x_1;b \bigg) \, \phi_T 
\bigg(\frac{|p_T-k_T|}{2},x_2;R-b \bigg) \, .
\label{ktfac}
\end{eqnarray}
with $\sigma_{s}$ as an effective interaction area which cancels with the $d^2b$ integration. In our notation $K_{kT}$ is a multiplicative constant, initially introduced 
to account for high-order corrections, to be adjusted according to the 
experimental data. $p_T$ is the transverse momentum of the produced gluon, 
$k_T$  and $p_T -k_T$ are the transverse momenta of the interacting gluons and $C_F=(N_c^2-1)/2N_c$ is the 
color factor. 

The Bjorken $x$ where the projectile and the target are probed can be written 
as a function of $p_T$ and $y$ according to the  $2\rightarrow1$ kinematics:
\begin{equation}
x_{1,2}=\frac{p_T}{\sqrt{s}}e^{\pm y}  \, .
\label{kinematics}
\end{equation}
The unintegrated gluon distribution function (UGD) gives the number of gluons 
per $x$, per transverse momentum unit. Following previous works \cite{rcbk,yuri_po,yuri_D,nara_26}, we use an UGD that is solution of the running coupling Balitisky-Kovchegov equation, which
is given by:

\begin{equation}
\varphi(k, x, \mathbf{R}) = \frac{C_F}{\alpha_s(k)(2\pi)^3} \int d^2  
\mathbf{r} \, e^{-i\mathbf{k}\cdot\mathbf{r}} \nabla_{\mathbf{r}}^2 
\mathcal{N}_A(r, x, \mathbf{R})\, .
\end{equation}
$\mathcal{N}_A$ is the dipole-hadron forward scattering amplitude for a gluon  
dipole (adjoint representation). This amplitude can be determined through the 
quark dipole scattering amplitude in the fundamental representation 
($\mathcal{N}_F$) according to:

\begin{equation}
\mathcal{N}_A(r, x, \mathbf{R}) = 2 \mathcal{N}_F(r, x, \mathbf{R}) - 
\mathcal{N}_F^2(r, x, \mathbf{R})\, ,
\end{equation} 
which is a solution to the running coupling  Balitsky-Kovchegov equation. We will use the MV-like initial conditions for the 
evolution

\begin{equation}
\mathcal{N}_F(r, x = x_0) = 1 - \exp               
\left[ -\frac{\left(r^2 Q_{s0}^2\right)^\gamma}{4} 
\ln \left( \frac{1}{\Lambda r} + e \right) \right] \,,
\end{equation}
with $Q_{s0}=0.157 \text{ GeV}$ as the initial saturation scale, 
$\gamma=1.101$ and $\Lambda=0.241 \text{ GeV}$.

\subsection{Hybrid formalism}

To compute the rapidity and transverse momentum distribution of final state 
particles in the forward-backward regions, we use the hybrid formalism \cite{rcbk,hibr}, where the parton production is given by:

\begin{equation}
\frac{dN}{dy \, d^2p_T \, d^2 \mathbf{{r}_{\perp}}} = 
\frac{K_{hib}}{(2\pi)^2} 
\left[
x_1 g(x_1, Q^2)\,\widetilde{\mathcal{N}}_A(p_T, x_2, Q^2) 
+ \sum_{q=u,d,s} x_1 q(x_1, Q^2)\,\widetilde{\mathcal{N}}_F(p_T, x_2, Q^2)
\right],
\label{hibrido}
\end{equation}
with $K_{hib}$ as the constant to be adjusted with the data, as in the       
$k_T$-factorization formalism. $g(x, Q^2)$ and $q(x, Q^2)$ are the gluon and  
quark parton distribution functions (PDFs) from the CT14 set \cite{ct14}. 
$\widetilde{\mathcal{N}}_{A,F}$, calculated at $Q^2 = \max \left( p_T^2,\; Q_s^2(x_2) \right)$, is given by:
\begin{equation}
\widetilde{\mathcal{N}}_{A,F}(p_T,x) = \int d^2r \, e^{i\vec{p}_T \cdot \vec{r}} [1 - \mathcal{N}_{A,F}(r,x)]
\end{equation}
The $x_{1,2}$ values are computed according to Eq.(\ref{kinematics}) in the  
forward region. For the backward production, otherwise, we rearrange the   
indices as $1 \rightarrow 2$, $2 \rightarrow 1$ to correctly match the UGD 
with the projectile and the PDF with the target.

\subsection{Intrinsic fluctuations}

Another ingredient that we include in our calculation are the intrinsic     
fluctuations. They are characterized by rare color charge fluctuations that 
occur in the projectile and the target, which, in the end, generate         
high-multiplicity collisions \cite{larry16,tgn26,tgn-d,yuri_po,ipjazzma,schenke,running_the_gamut}. Numerically, these fluctuations 
are modeled by a log-normal distribution of parameter $\sigma$
\begin{equation}
P\left(\ln\left(\frac{Q_s^2}{ Q_0^2 }\right)\right) = 
\frac{1}{\sqrt{2\pi\sigma}} 
\exp\left(-\frac{\ln^2\left(Q_s^2 /  Q_0^2 \right)}{2\sigma^2}\right) \, ,
\end{equation}
and indicate the probability of the saturation scale to fluctuate around its    
value ($Q_0$). The peak of this distribution occurs at $(Q_s/Q_0)^2 \approx 1$, 
implying that it is very likely to have no fluctuations. However it is 
possible to have rare $(Q_s/Q_0)^2 \gg 1$ deviations. In \cite{tgn26} we used $0.7<\sigma<0.8$ in pp simulations with Gaussian initial conditions and KLN UGD (Karzeev-Levin-Nardi), but, since we are using a different UGD, it will be updated according to the experimental data.

\subsection{Simulations}

\begin{figure}[!h]
\centering
\begin{tabular}{ccc} 
\includegraphics[scale=0.35]{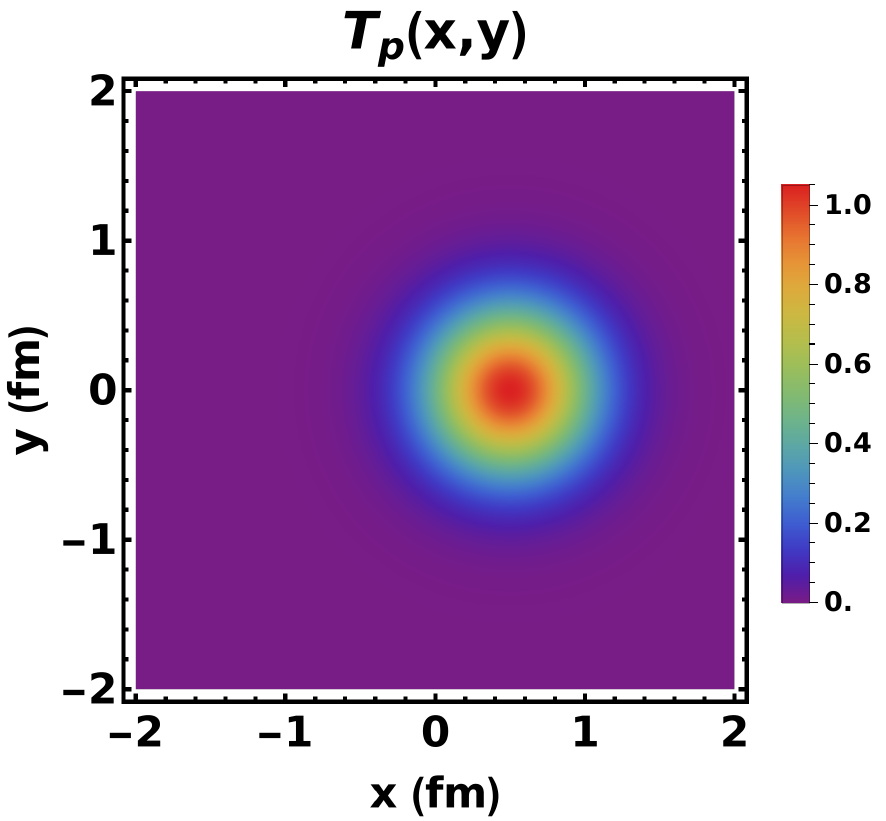}& 
\includegraphics[scale=0.35]{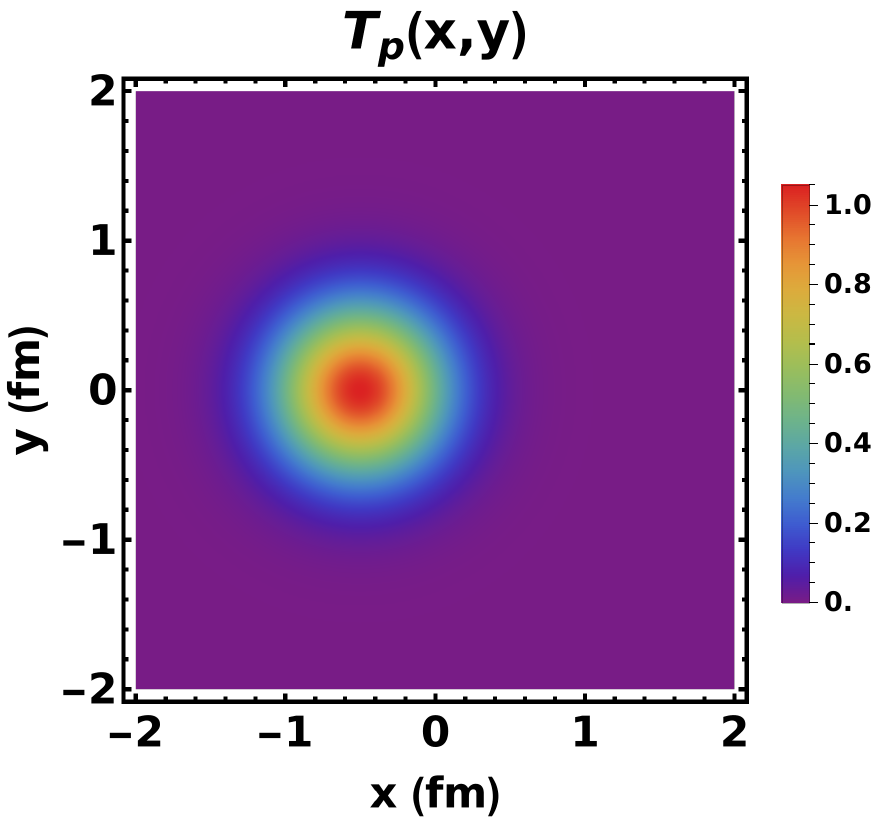}& 
\includegraphics[scale=0.35]{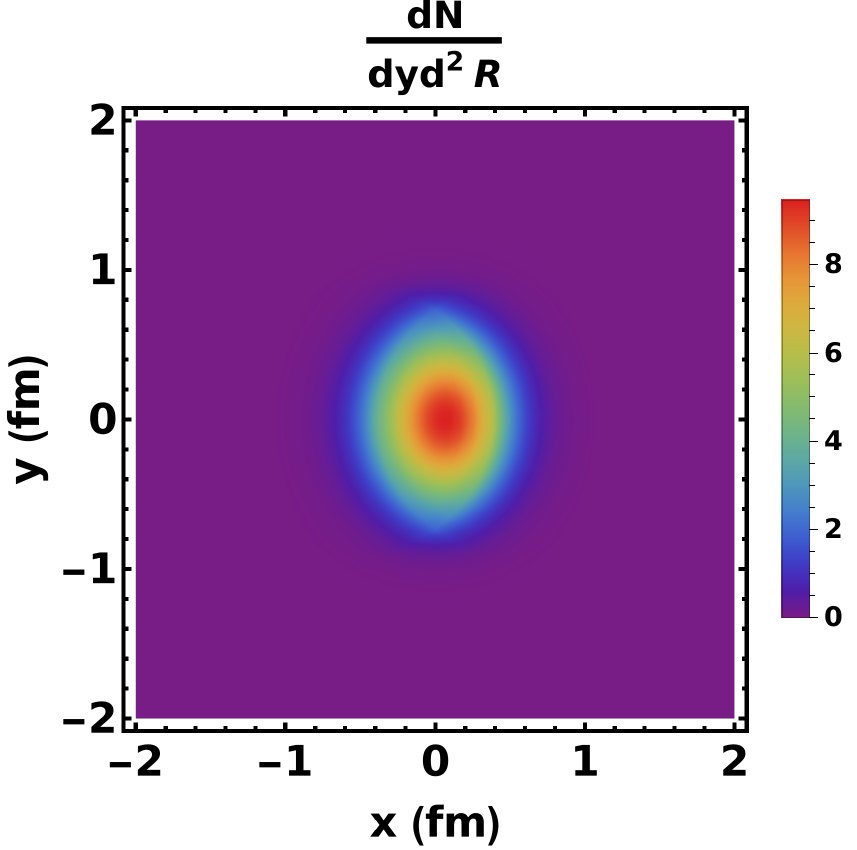}\\
(a) & (b) & (c)\\
\end{tabular}
\caption{a) Projectile  and b) target proton thickness.          
c) $dN/dy \,  d^2 \mathbf{r}_{\perp} $ of produced partons at midrapidity. 
Adapted from \cite{tgn26}.}
\label{simualtion}
\end{figure}

We simulated pp collisions in a Monte Carlo event generator which combines    
Gaussian proton initial conditions, $k_T$-factorization and hybrid formalism. 
The proton thickness (z-integrated density) is given by:
\begin{equation}
T_p(\mathbf{r}_\perp)=\frac{e^{-\frac{\mathbf{r}_\perp^2}{2B_p}}}{2\pi B_p}\, ,
    \label{thicknessgauss}
\end{equation}
with $B_p \approx 3.8 \text{ GeV}^2$, estimated through comparisons with 
HERA measurements \cite{schenke,bp_raju}. The simulator calculates the parton  
production for a given rapidity, in the transverse plane, in the overlap 
region. An example of a simulated event 
	is shown in Fig \ref{simualtion}.  

To obtain the pseudorapidity distribution, we perform  a $y\rightarrow \eta$ 
variable transformation by using the jacobian  
${\cosh \eta}/{\sqrt{\cosh^2 \eta + m^2 / k^2}}$                          
with $m=0.35\text{ GeV}$, and then integrate over $d^2\mathbf{r}_{\perp}$ 
in each collision. 
After that, we assume the parton-hadron duality to convert partons into      
charged particles. According to this assumption, $N_{ch} \propto N_{partons}$, 
with the proportionality absorbed in $K_{kT}$ and in $K_{hib}$ in        
Eq.(\ref{ktfac}) and  Eq.(\ref{hibrido}), respectively.

\subsection{The matching procedure}

\begin{figure}[!h]
\centering
\begin{tabular}{cc} 
\includegraphics[scale=0.38]{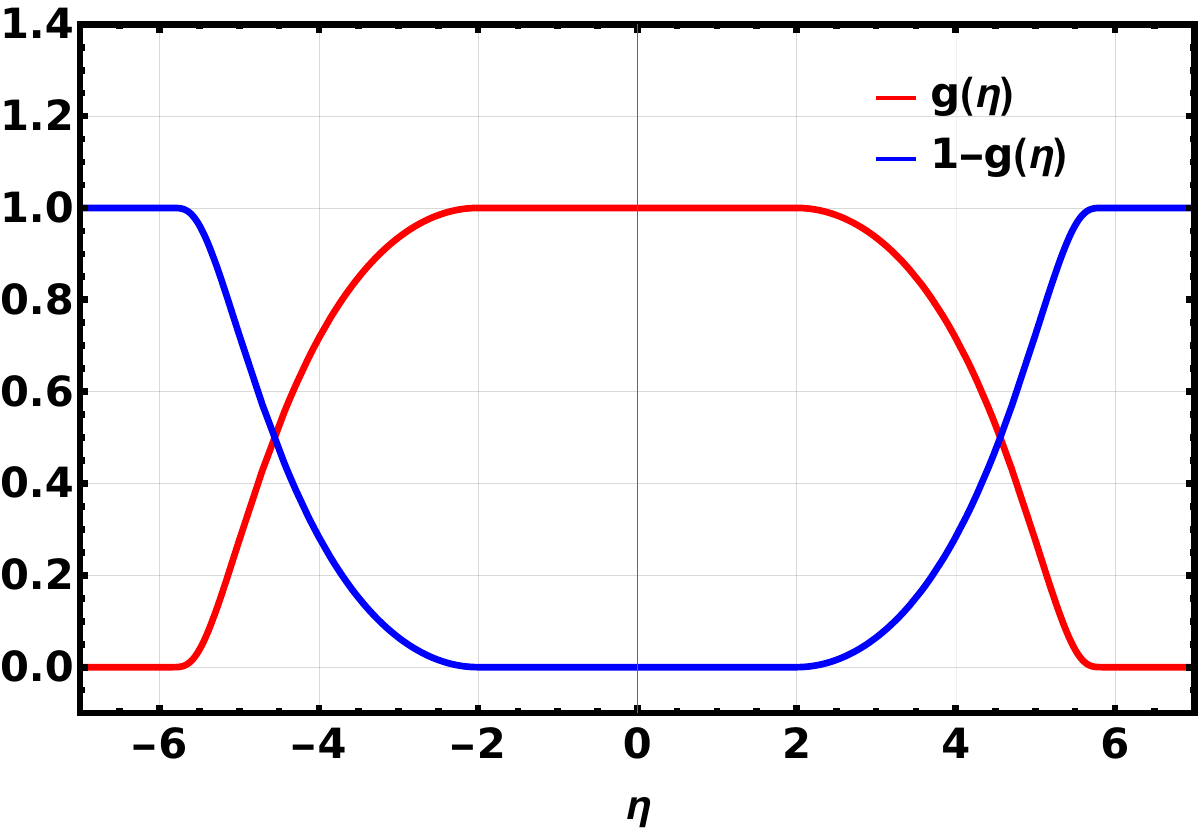}& 
\includegraphics[scale=0.4]{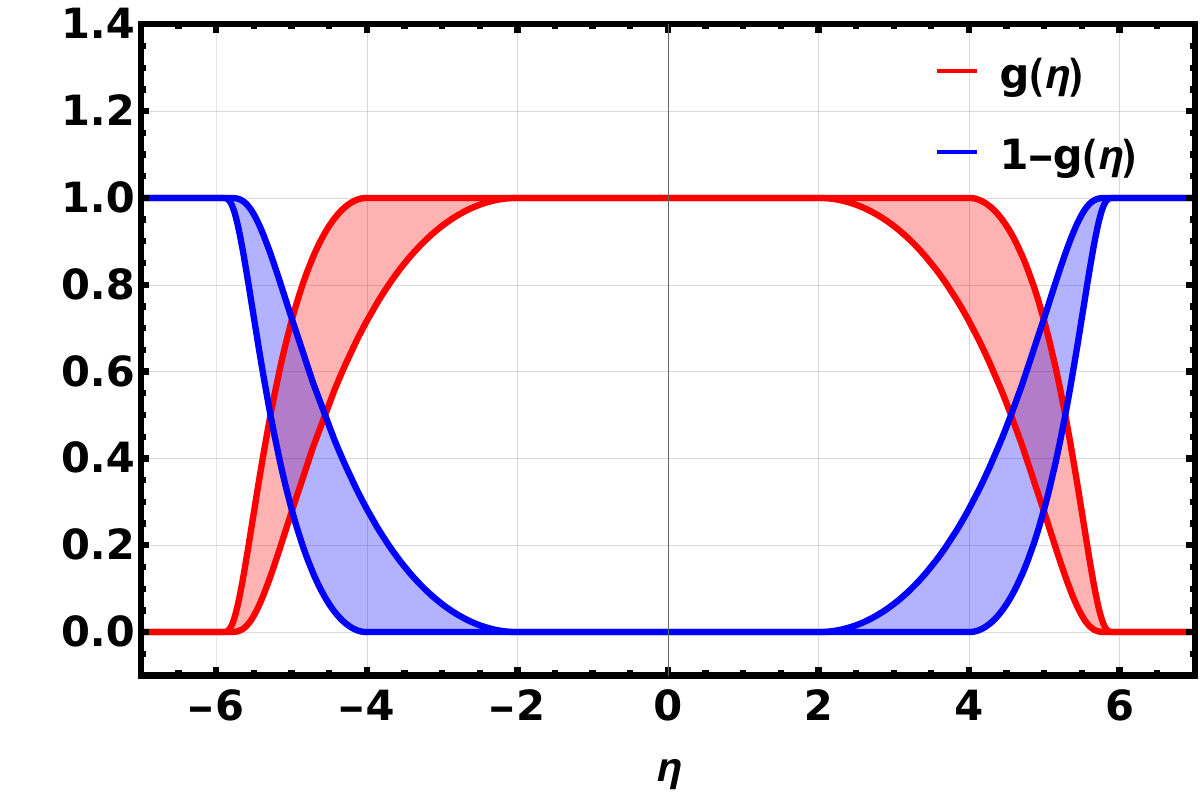}\\
(a) & (b) \\
\end{tabular}
\caption{a) Weight function with the parameters of Set I. b) 
Weight function with the parameters of Set I and Set II and bands between the 
curves.}
\label{g}
\end{figure}

As discussed before there is not a sharp separation point between  
$k_T$-factorization and hybrid formalism, so the transition over 
each approach is better assumed to be over a given region of 
(pseudo-)rapidity. In this work we propose the 
inclusion of the following $\eta$-dependent weight function 
to implement the transition between the two formalisms:
\begin{equation}
\frac{dN}{d\eta}(\eta) = g(\eta) \, \frac{dN_{kT}}{d\eta}(\eta) 
+ [1 - g(\eta)] \, \frac{dN_{hib}}{d\eta}(\eta) \, ,
\label{matching}
\end{equation}
with 
\begin{equation}
g(\eta) =
\begin{cases} 
1 & |\eta| \leq \eta_0 \\ 
\exp \left( 1 - \frac{1}{1 - \left( \frac{|\eta| - \eta_0}{\delta} \right)^2} 
\right) & \eta_0 < |\eta| < \eta_0 + \delta \\ 
0 & |\eta| \geq \eta_0 + \delta 
\end{cases} \, .
\end{equation}
In this notation $\eta_0$ is the pseudorapidity where the transition starts    
and $\delta$ is the transition window. The plot for $\eta_0=2$ and $\delta=4$ is 
shown in Fig.\ref{g}a. In Fig.\ref{g}b we show the same plot with filled bands between the previous curve and the one for $\eta_0=4$ and $\delta=2$. The resulting interpolating functions will be used to estimate the uncertainty regarding the transition between the two components of the model.

After defining the matching procedure, the pseudorapidity    
distribution of Eq. (\ref{matching}) can be integrated in the appropriate 
$\eta$ domain in an event-by-event basis, yielding $N_{ch}$ and 
the multiplicity distribution.

\section{Results and discussion}

\subsection{Pseudorapidity distributions}

\begin{figure}[!h]
\centering
\begin{tabular}{cc} 
\includegraphics[scale=0.3]{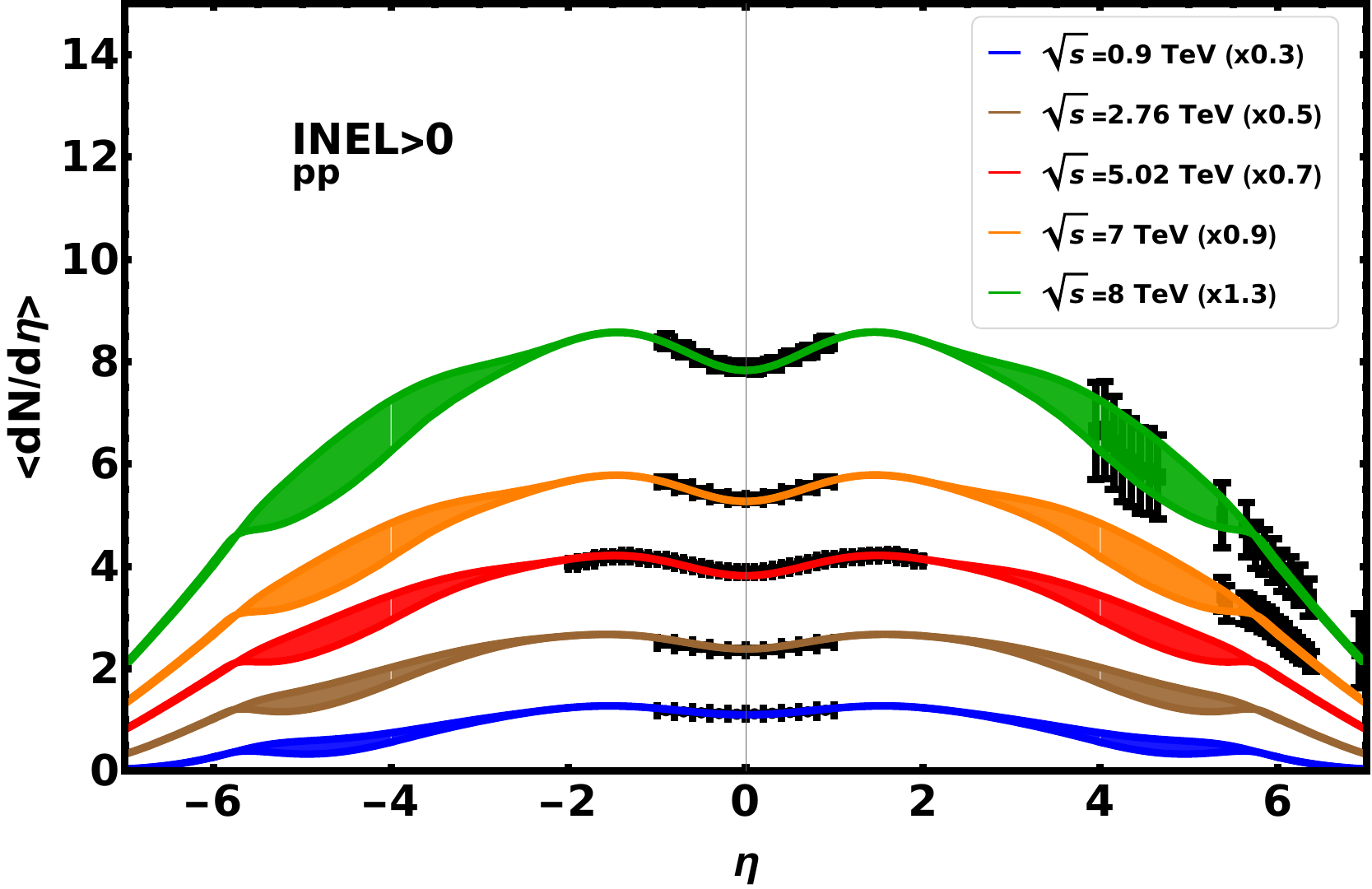}& 
\includegraphics[scale=0.3]{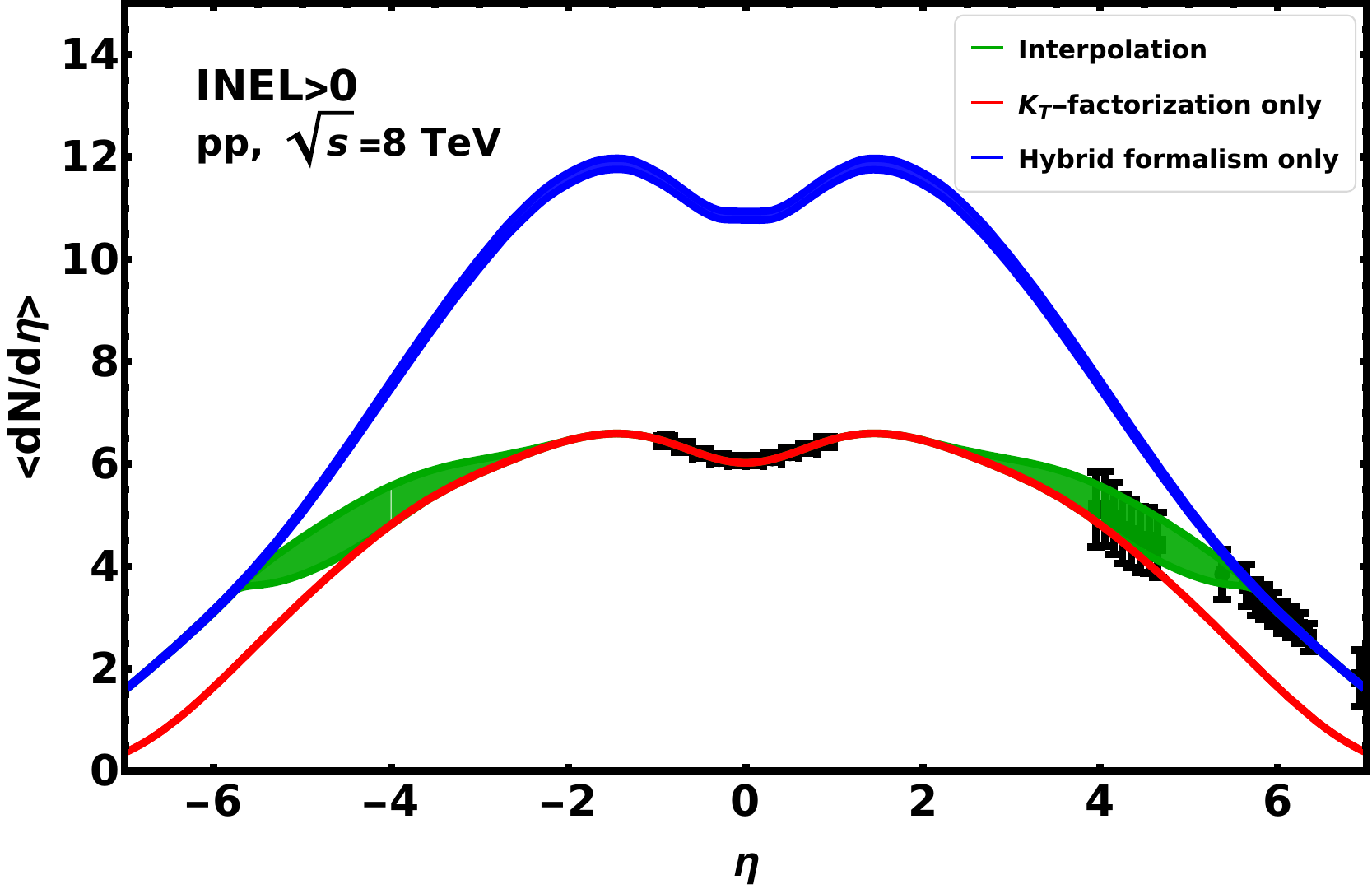}\\
(a) & (b) \\
\end{tabular}
\caption{a) Global fit of $\langle dN/d\eta \rangle$ obtained with 
Eqs. (\ref{ktfac}, \ref{hibrido}, \ref{matching}).                     
b) Results for $\sqrt{s}=8\text{ TeV}$. The experimental data are from 
\cite{ALICE_5,ALICE_17,TOTEM7,TOTEM8}.}
\label{dndeta}
\end{figure}

There are five parameters in the model: 
$\eta_0$, $\delta$, $\sigma$, $K_{kT}$ and $K_{hib}$. We first choose two sets of $\eta_0$,        
$\delta$ and $\sigma$ in order to estimate the uncertainties, and then fix $K_{kT}$ and $K_{hib}$ through a global fit  
of the measured pseudorapidity distributions for each set. We consider data taken at 
center of mass energies from $0.9$ TeV to $8$ TeV, as shown in 
Fig. \ref{dndeta}a. They refer to inelastic events with at least one particle
produced in the corresponding pseudorapidity interval and are labeled INEL$>$0. 
The resulting parameters for each combination of $\eta_0$ and $\delta$ are 
presented in Table \ref{parameters}.
\begin{table}[h!]
\centering
\caption{Resulting parameters needed in Eq.(\ref{matching}).}
\begin{tabular}{l c c c c c c}
\hline
& $\eta_0$ & $\delta$ & $\sigma$ & $K_{\mathrm{kt}}$ & $K_{\mathrm{hib}}$ 
& $\chi^2_{\mathrm{red}}$ \\
\hline
Set 1 & 2    & 4    & 0.5  & 9 & 5.6 & 0.65 \\
Set 2 & 4    & 2    & 0.4  & 8.7 & 5.6 & 0.6 \\
\hline
\hline
\end{tabular}
\label{parameters}
\end{table}

In Fig. \ref{dndeta}b we show the resulting fit of the $\sqrt{s}=8\text{ TeV}$ 
	data as well as each individual component in Eq.(\ref{matching}).    
It can be seen that the $k_T$-factorization is able to explain the midrapidity  
data but, as the pseudorapidity increases, it starts to underestimate them.  
On the other hand, the hybrid formalism describes correctly the very forward   
data, but badly overestimates the data in the central region. 
These results suggest that there is a transition region where both formalisms
contribute partially. It is represented by the ``interpolation" curve, calculated with Eq.(\ref{matching}). 

\begin{figure}[!h]
\centering
\begin{tabular}{c} 
\includegraphics[scale=0.4]{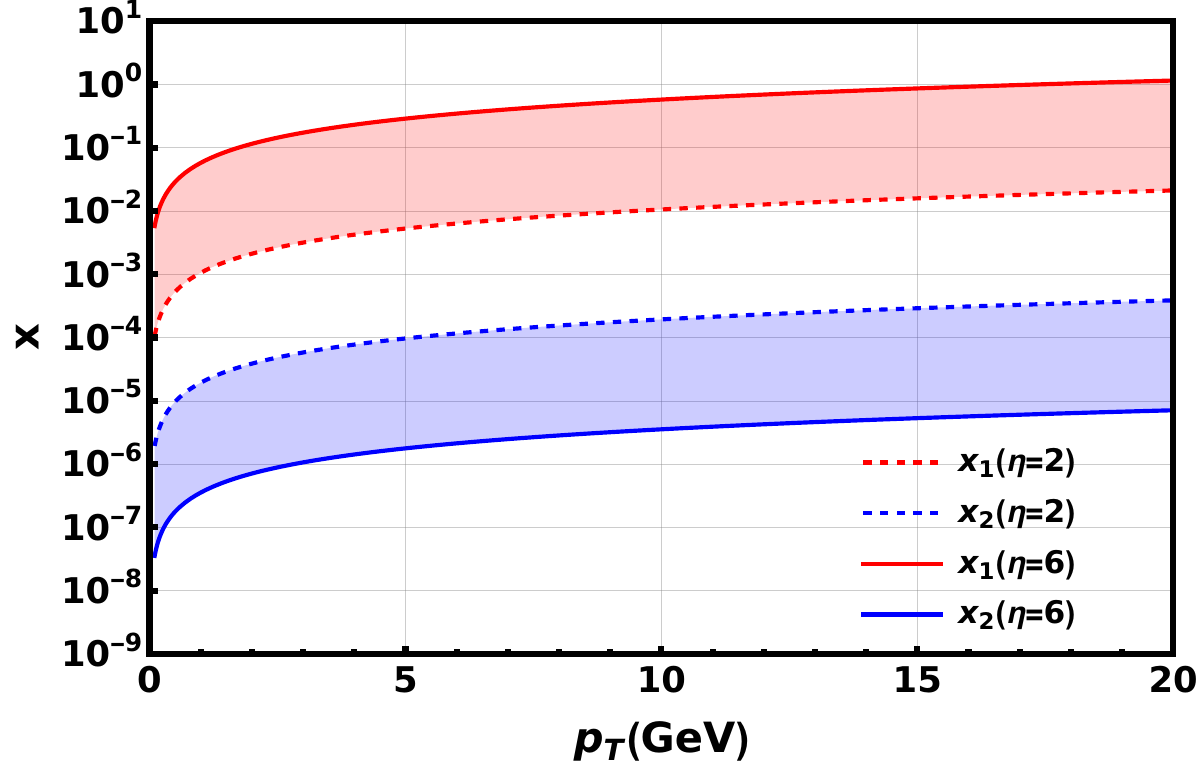}\\
\end{tabular}
\caption{Phase space plot for the transition region.}
\label{phase}
\end{figure}

The $(p_T,x)$  region where the transition occurs can be estimated from the 
parameters listed in Table \ref{parameters} and the process kinematics. 
For the wider transition window (set 1), the $\eta$ boundaries can be inserted 
in Eq.(\ref{kinematics}), for a given $\sqrt{s}$, and the curves of $x_1,x_2$ 
as a function of $p_T$ can be plotted. For $\eta=2$ and $\sqrt{s}= 7 \text{ TeV}$, 
the curves are represented by the dashed lines in Fig. \ref{phase}. The same curves for  
$\eta=6$ are represented by the solid lines. The filled regions represent    
the phase space points where the transition between the $k_T$-factorization  
and hybrid formalism occurs (a similar analysis can be made for the backward 
pseudorapidity region by changing the $x_i$ indices).

By repeating the same procedure, but now considering the experimental value for 
the average transverse momentum of produced particles in proton-proton collisions 
collisions at 7 TeV measured by the CMS collaboration ($\langle p_T \rangle=0.545 \pm 0.005 \text{(stat.)} \pm \text{0.015 (syst.))} \text{GeV/c}$)\cite{mean_pt}, it possible 
to make qualitative estimates for transition boundaries in the ($x_1$,$x_2$) plane.
For $\eta=2$, the smaller pseudorapidity where the transition starts in our model we 
have, $(x_1,x_2)\approx(6\times 10^{-4},10^{-5})$. For  $\eta=6$, the pseudorapidity 
where the hybrid formalism starts to be the unique formalism, 
$(x_1,x_2)\approx(3\times 10^{-2},2\times 10^{-7})$.

For $\sqrt{s}=5.02 \text{ TeV}$, the same qualitative estimations can be made with a more recent mean $p_T$ data from LHCb \cite{LHCb_pt}. For $\eta=2$, $(x_1,x_2)\approx( 10^{-3},2 \times 10^{-5})$, and For $\eta=6$, $(x_1,x_2)\approx(7 \times 10^{-2}, 4 \times 10^{-7})$.

\subsection{Multiplicity distributions and KNO scaling}

\begin{figure}[!h]
\centering
\begin{tabular}{cc} 
\includegraphics[scale=0.3]{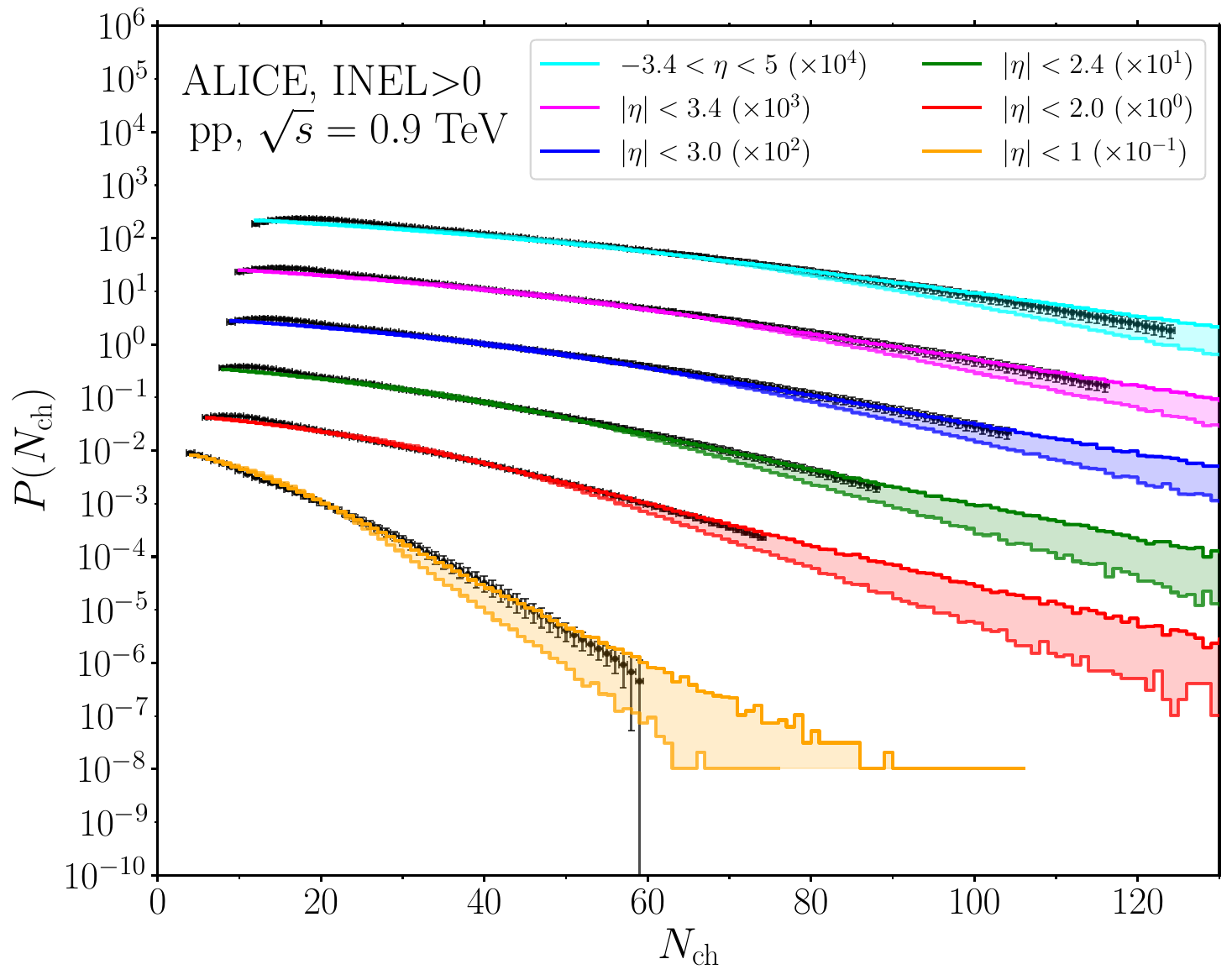}& 
\includegraphics[scale=0.3]{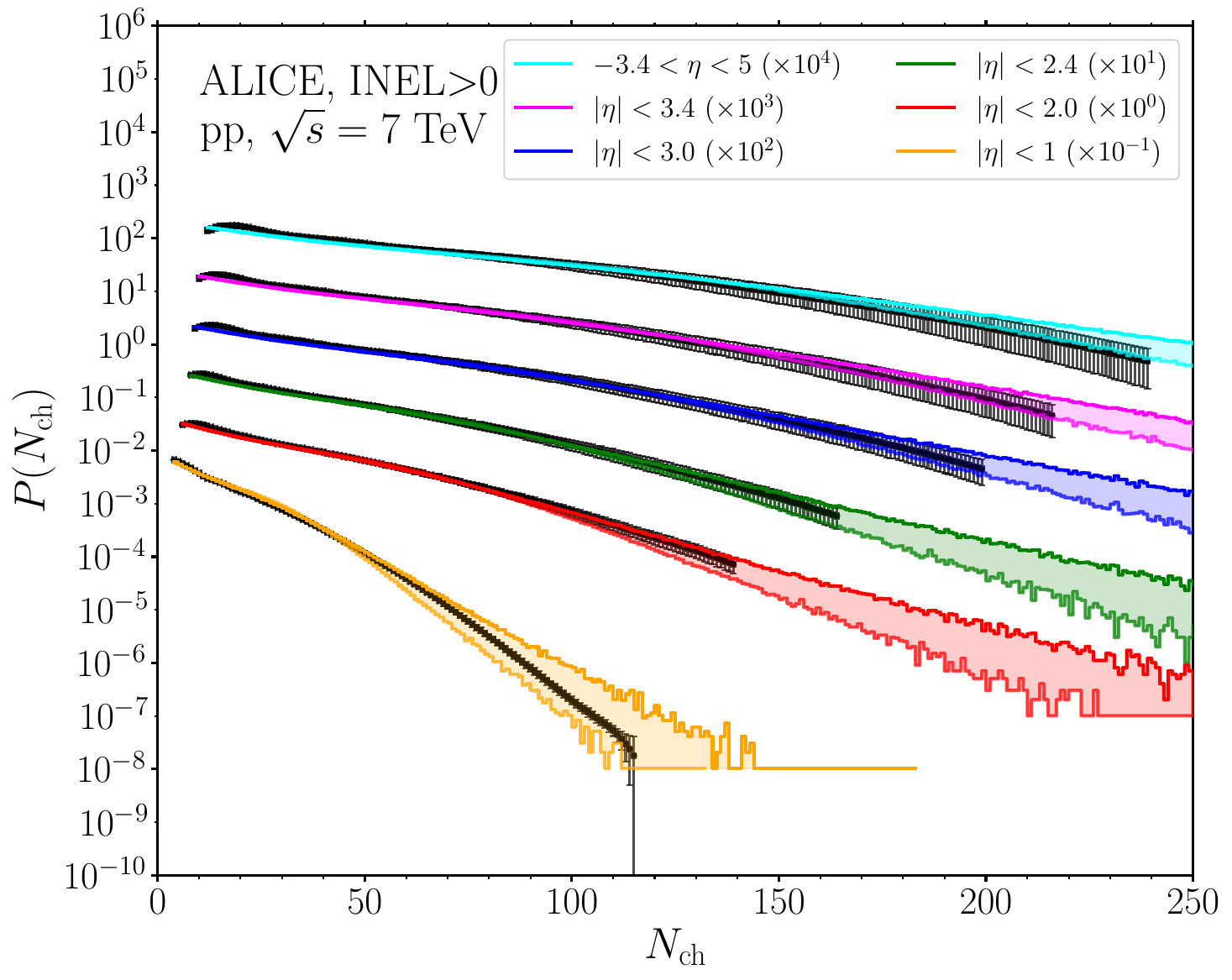}\\
(a) & (b) \\
\includegraphics[scale=0.3]{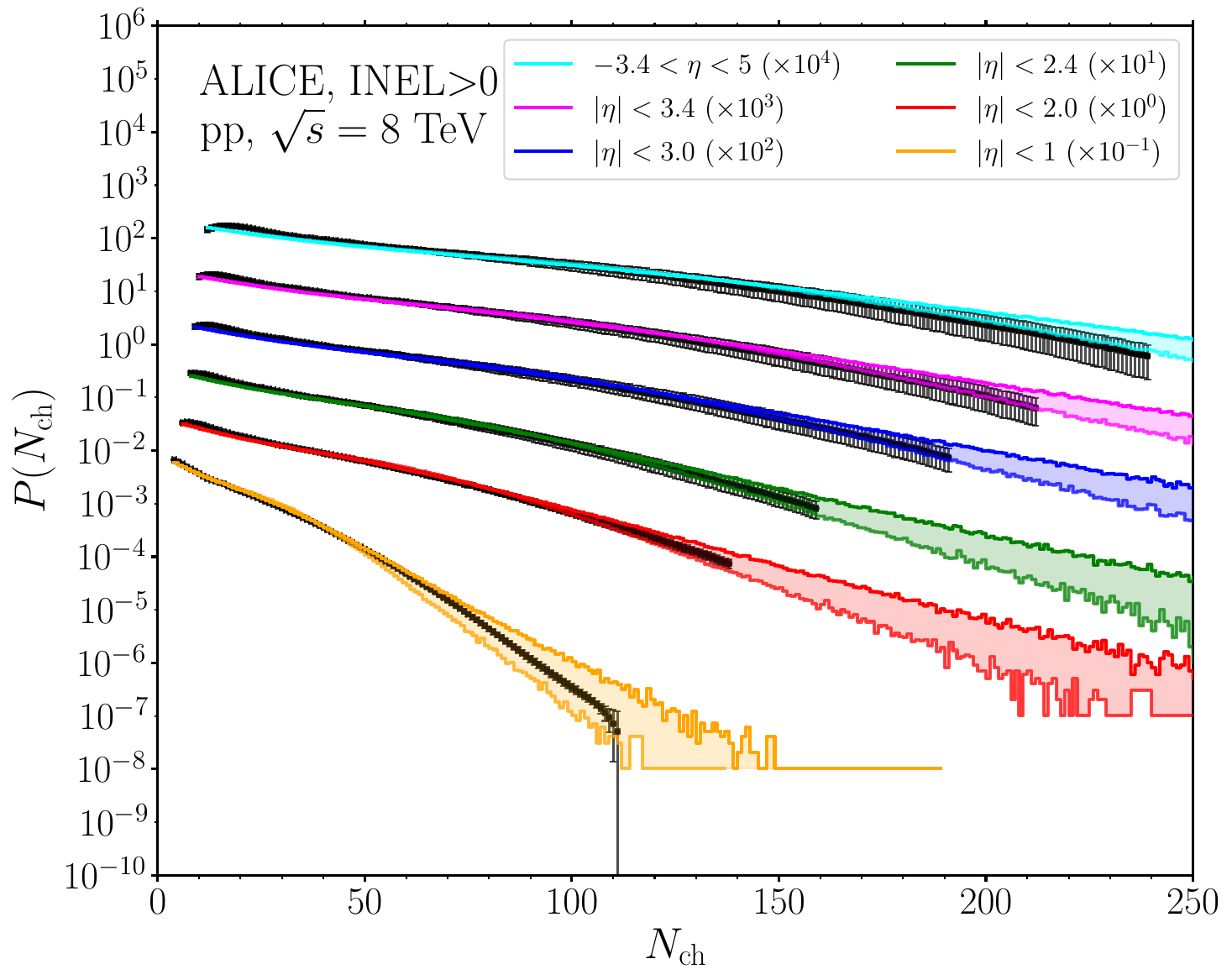}& 
\\
(c) &  \\
\end{tabular}
\caption{Multiplicity distributions for a) $\sqrt{s}=0.9$ TeV, b) $\sqrt{s}=7$ TeV and c) $\sqrt{s}=8$ TeV. The theoretical uncertainties were estimated through the sets of table \ref{parameters}. The experimental data are from \cite{ALICE_17,ALICE_wide}.}
\label{multiplicity}
\end{figure}

After fixing the parameters, we can calculate the multiplicity distributions.  
The results for INEL$>$0 data at $\sqrt{s}=0.9,\, 7,\, 8 \text{ TeV}$ are shown 
in Fig. \ref{multiplicity}. We can reproduce well all the data, 
from narrow pseudorapidity windows near $\eta =0$ to wider windows. The band widths were estimated through the sets from table \ref{parameters}.

\begin{figure}[!h]
\centering
\begin{tabular}{ccc} 
\includegraphics[scale=0.3]{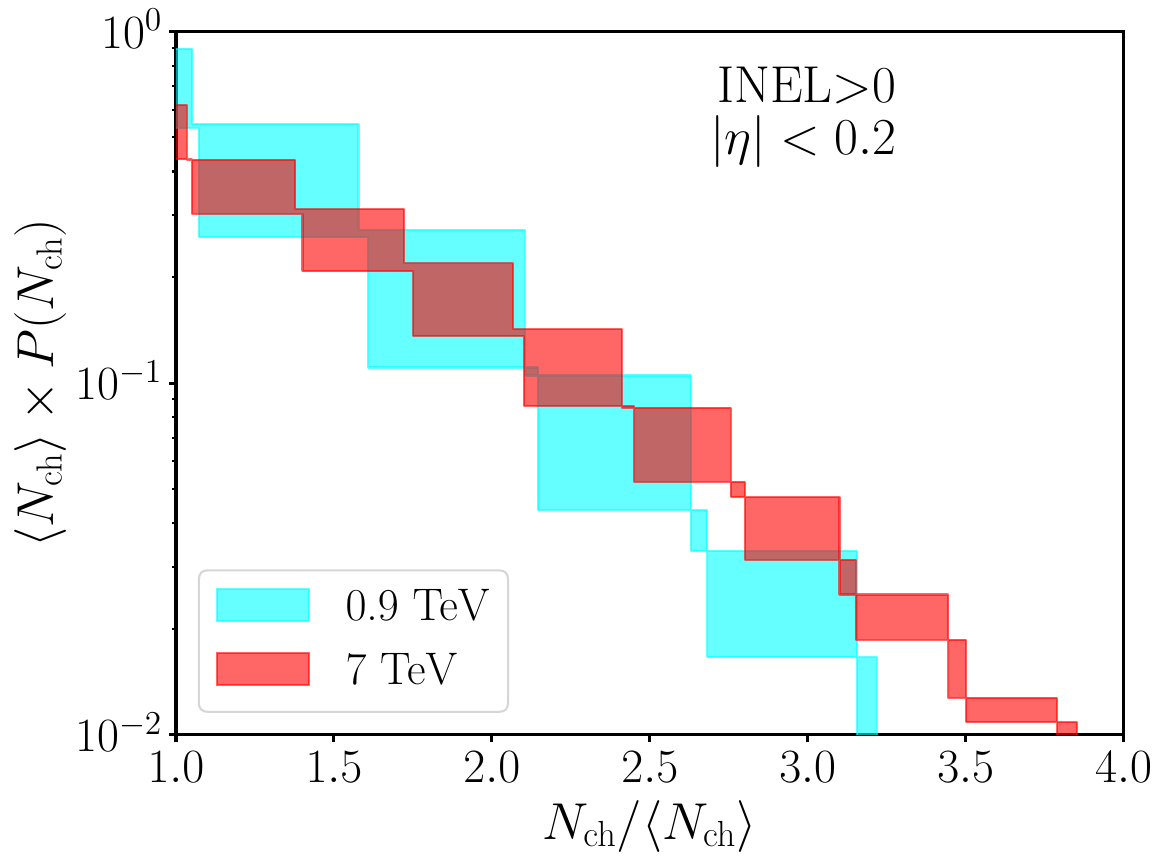}& 
\includegraphics[scale=0.3]{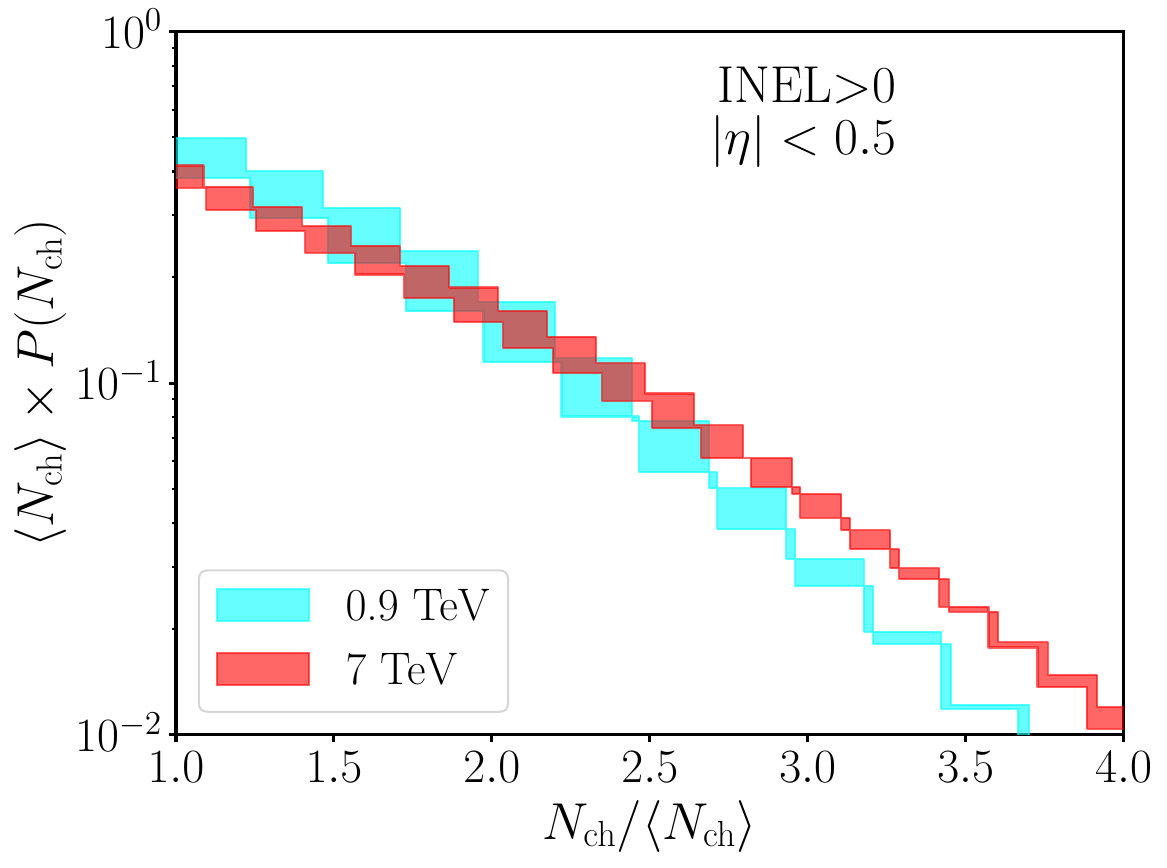} & \includegraphics[scale=0.3]{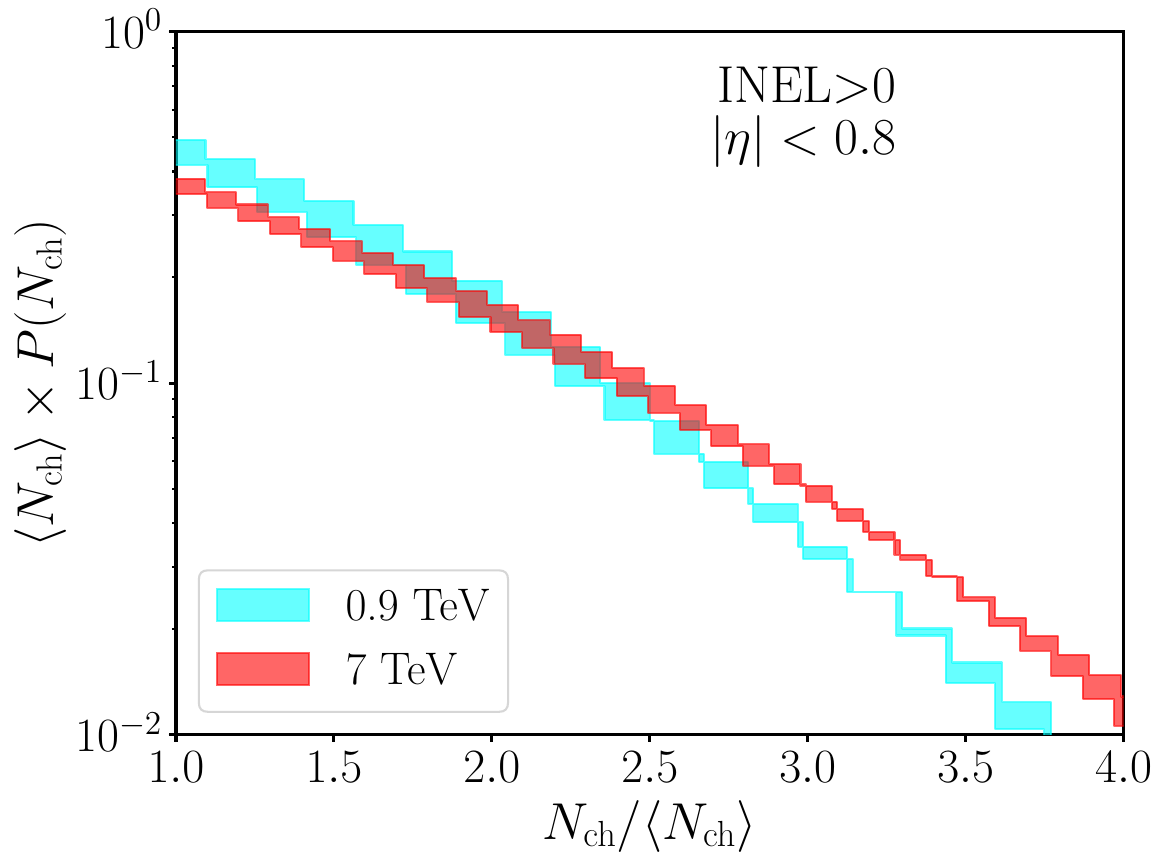}\\
(a) & (b) & (c)\\

\includegraphics[scale=0.3]{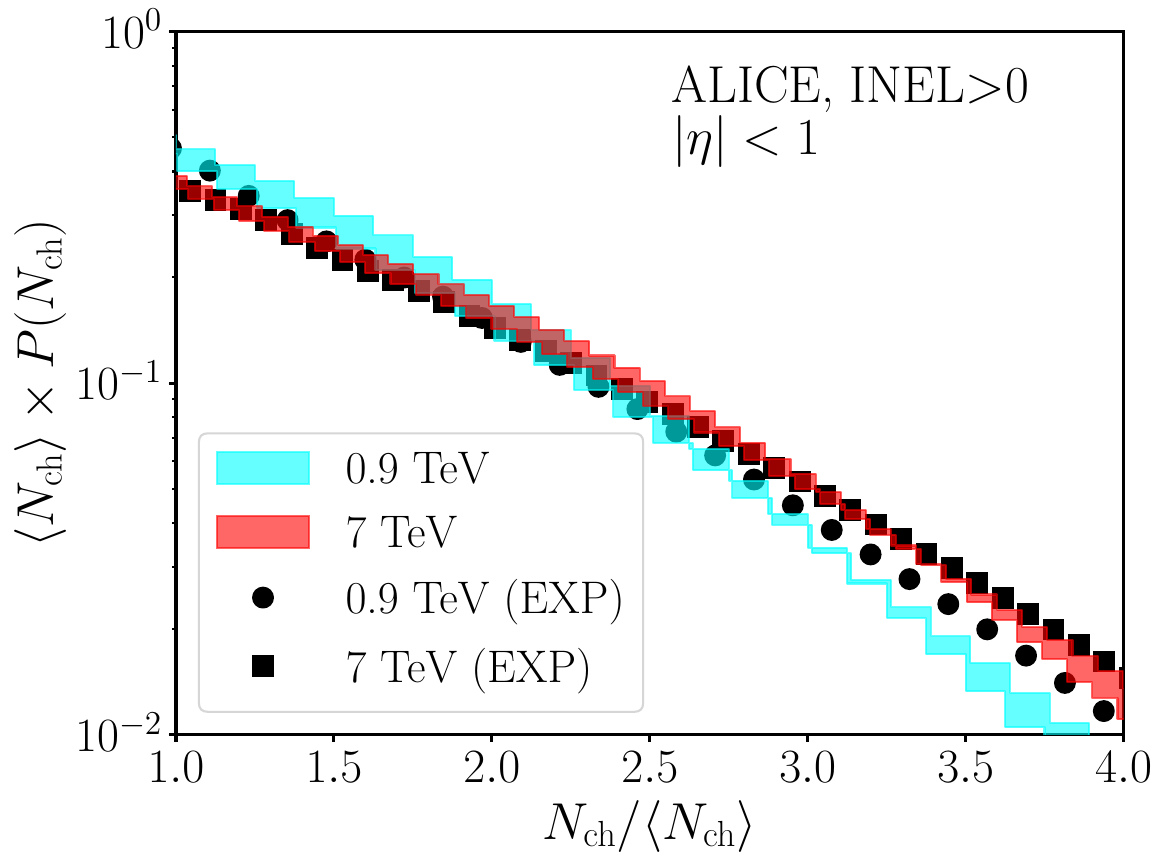}& 
\includegraphics[scale=0.3]{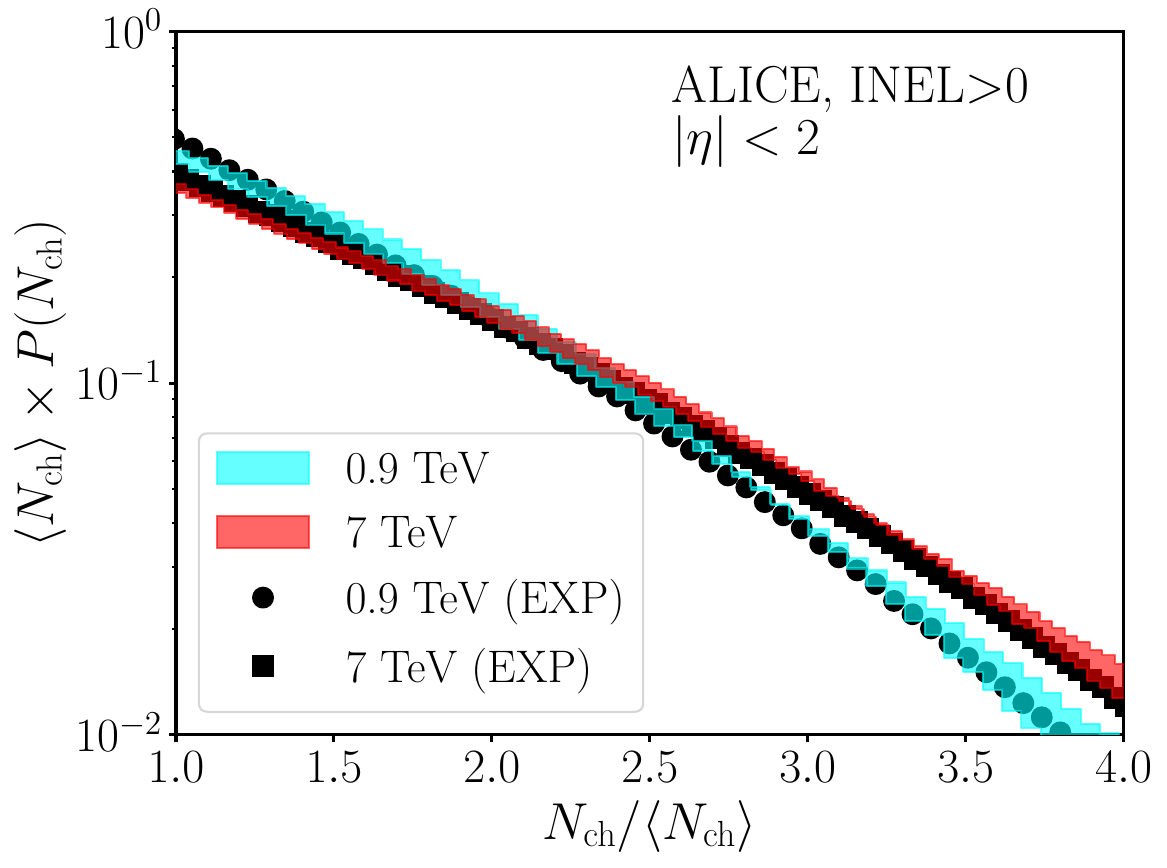} & \includegraphics[scale=0.3]{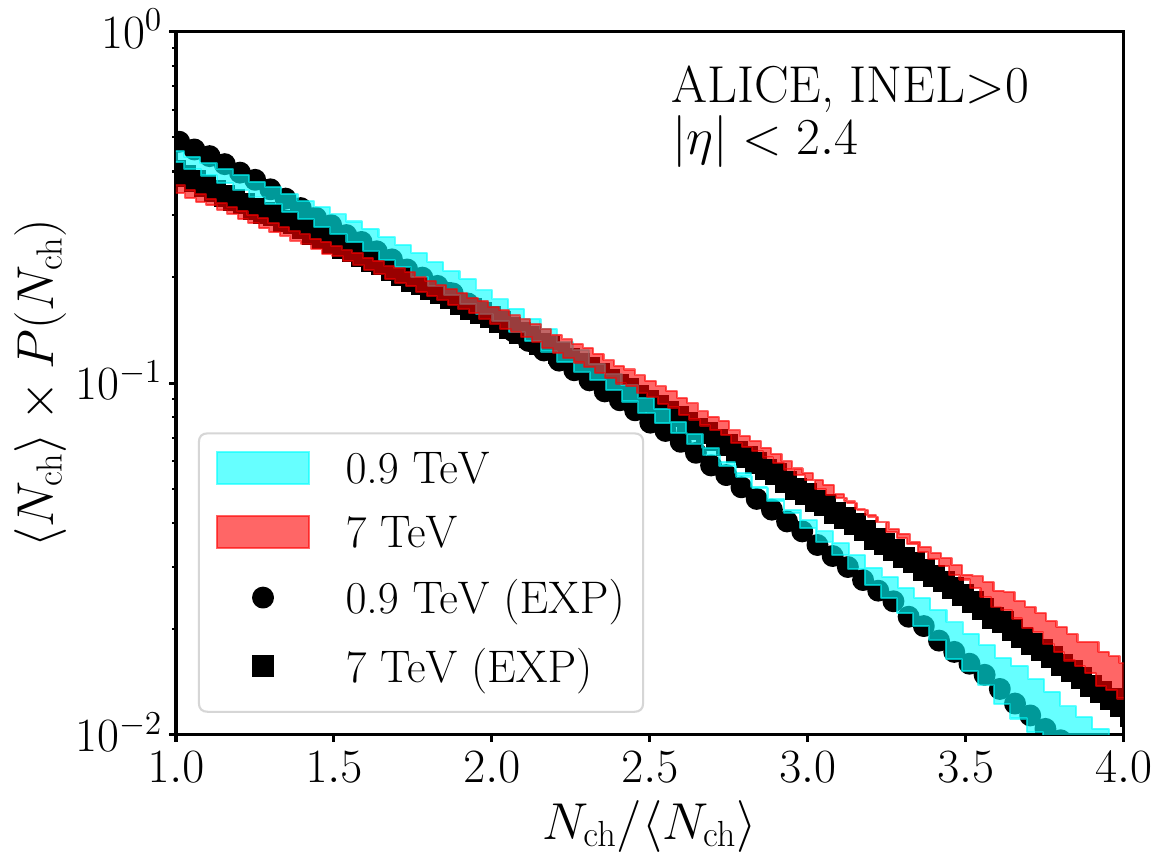}\\
(d) & (e) & (f)\\

\includegraphics[scale=0.3]{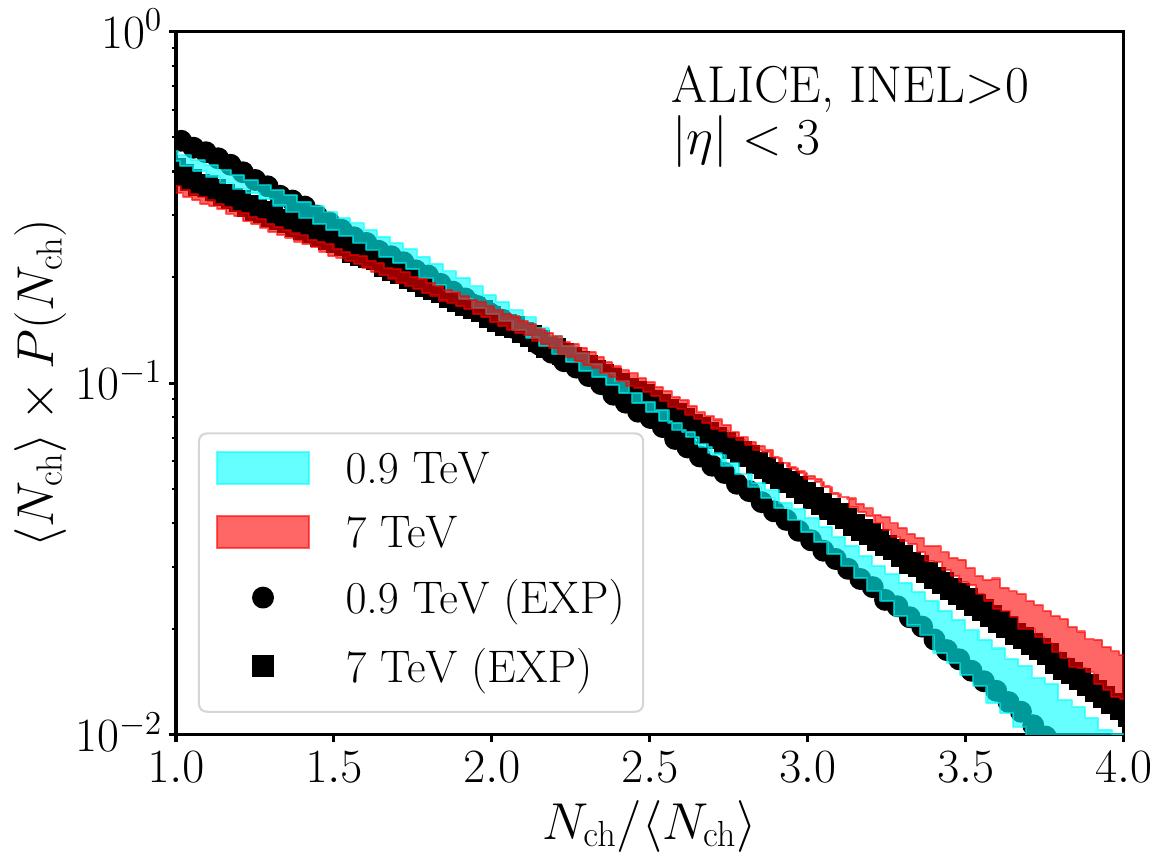}& 
\includegraphics[scale=0.3]{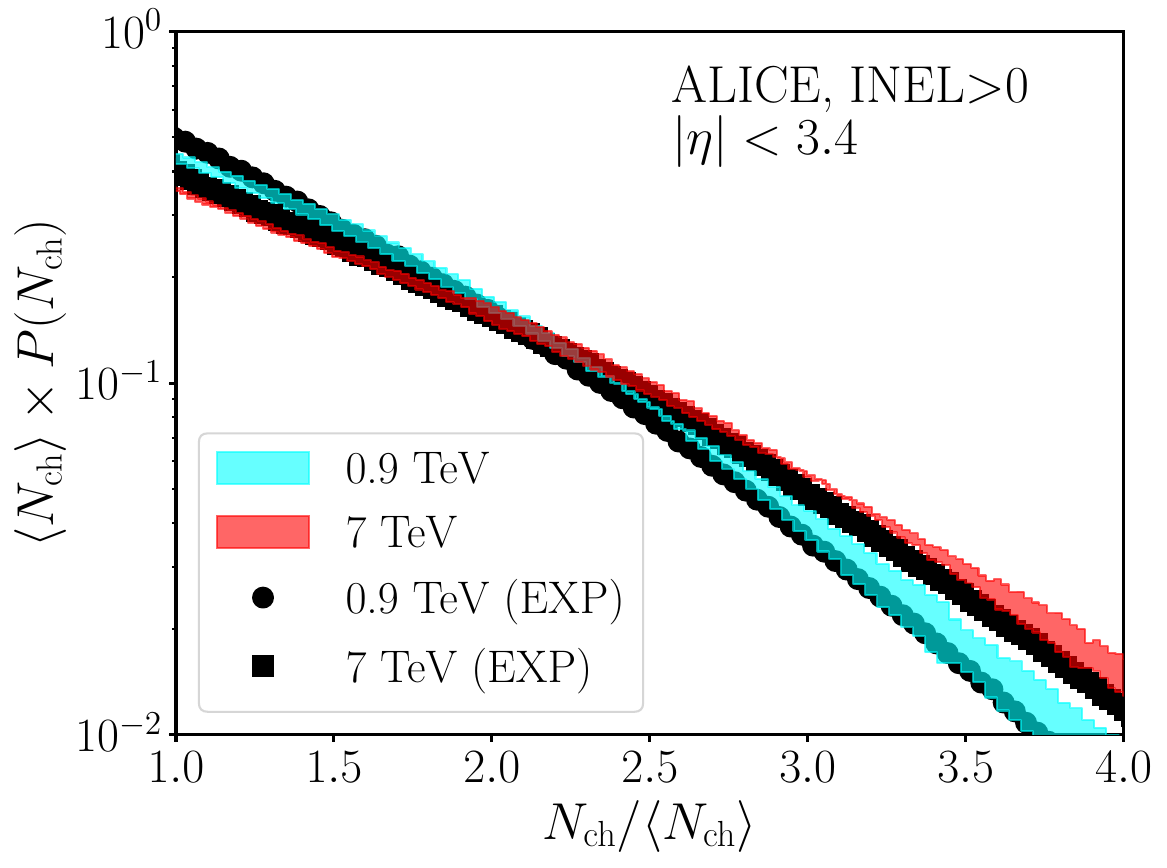} & \includegraphics[scale=0.3]{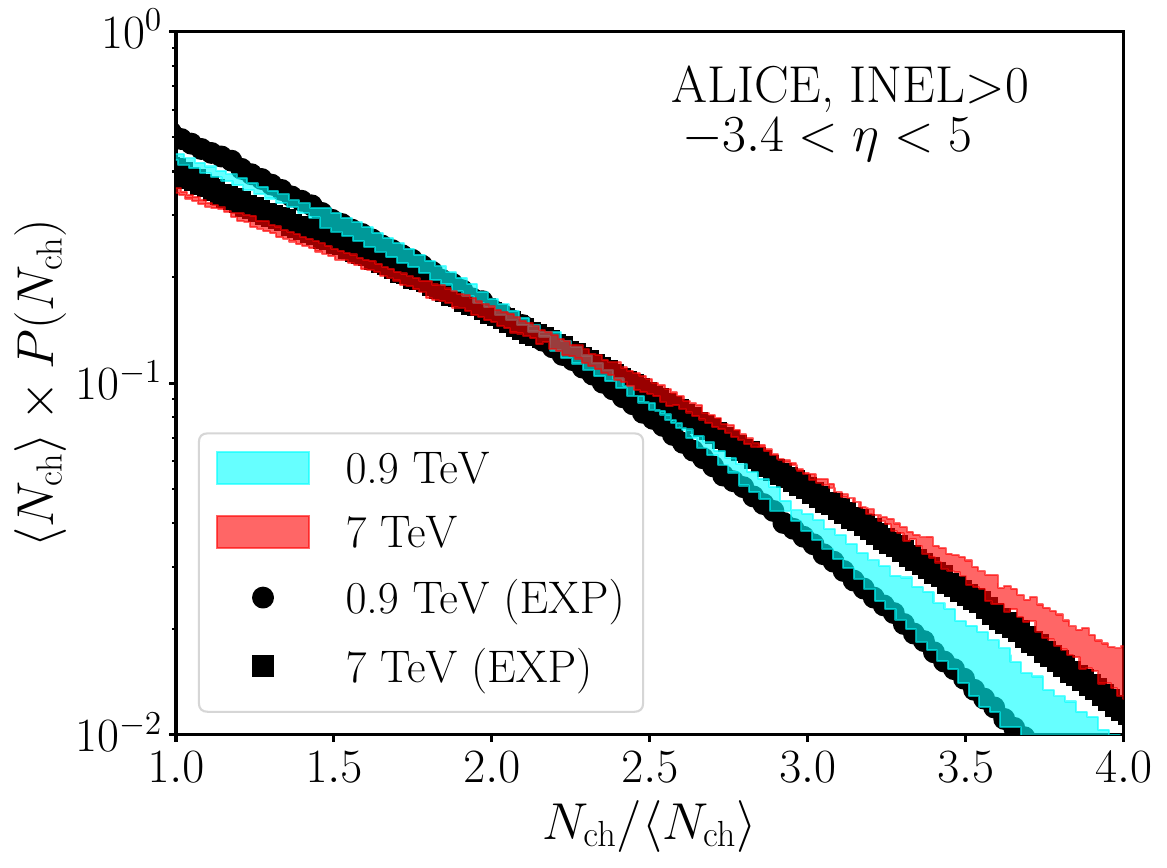}\\
(g) & (h) & (i)\\
\end{tabular}
\caption{KNO plot for a) $|\eta|<0.2$, b) $|\eta|<0.5$, c) $|\eta|<0.8$, d) $|\eta|<1$, e) $|\eta|<2$, f) $|\eta|<2.4$, g) $|\eta|<3$, h) $|\eta|<3.4$ and i) $-3.4<\eta<5$. The experimental data are from \cite{ALICE_17,ALICE_wide}.}
\label{kno}
\end{figure}

In Fig. \ref{kno} we present the results in the KNO form. From the figure, it 
can be seen that KNO scaling is observed for 
small $|\eta|$ and is violated as $|\eta|$ increases. All the comparisons are 
made in the intermediate multiplicity region ($1<z<4$). 
For $z<1$ the production is dominated by diffractive processes, not accounted 
for by our formalism. For $z\approx 4$ the theoretical results start to        
overestimate the experimental data. This behavior was already found in our  
previous work \cite{tgn26} for Gaussian initial conditions. 

\begin{figure}[!h]
\centering
\begin{tabular}{cc} 
\includegraphics[scale=0.4]{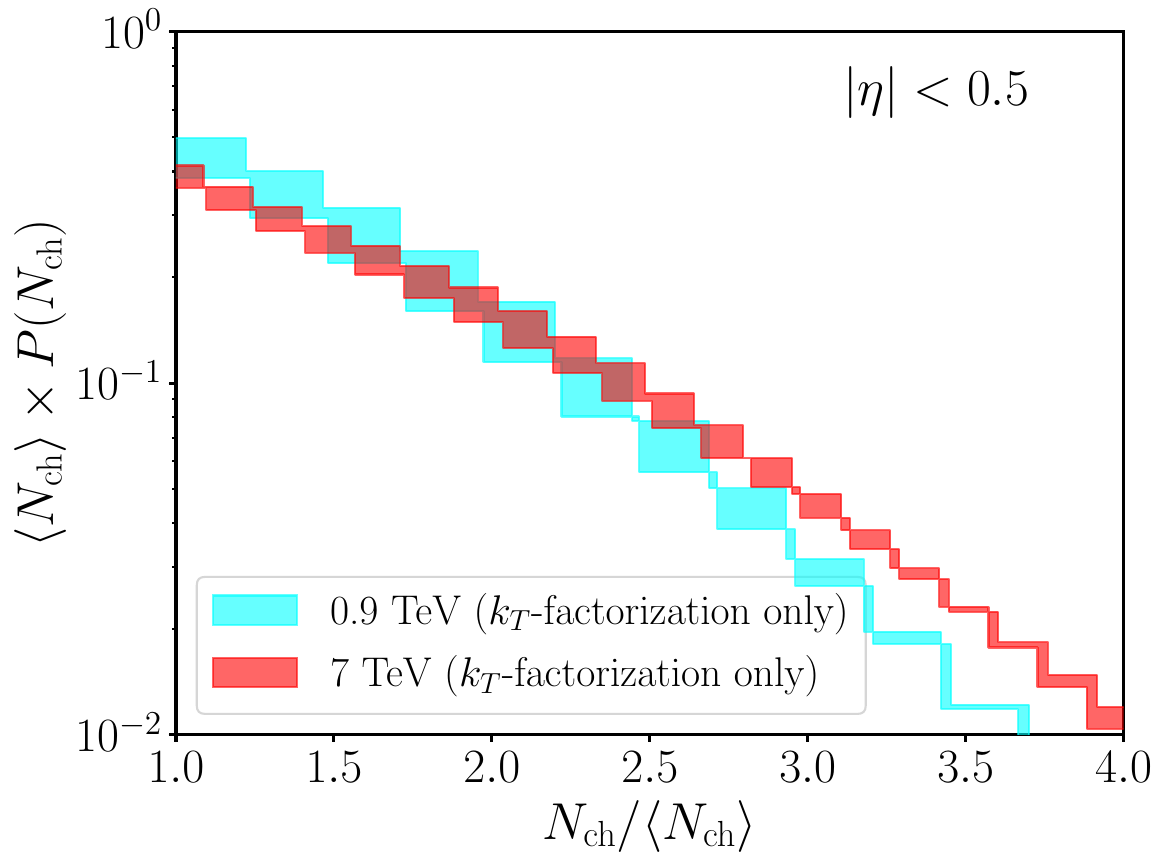}& 
\includegraphics[scale=0.4]{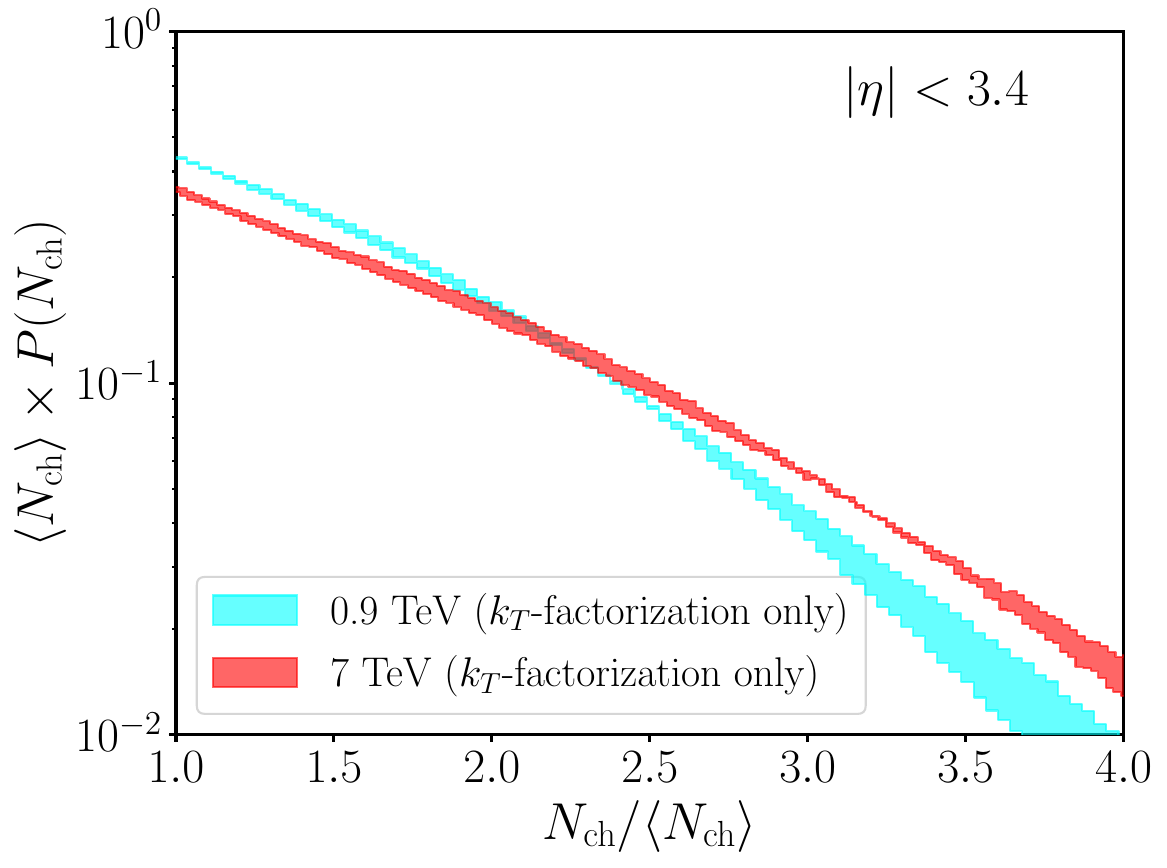} \\
(a) & (b) \\

\includegraphics[scale=0.4]{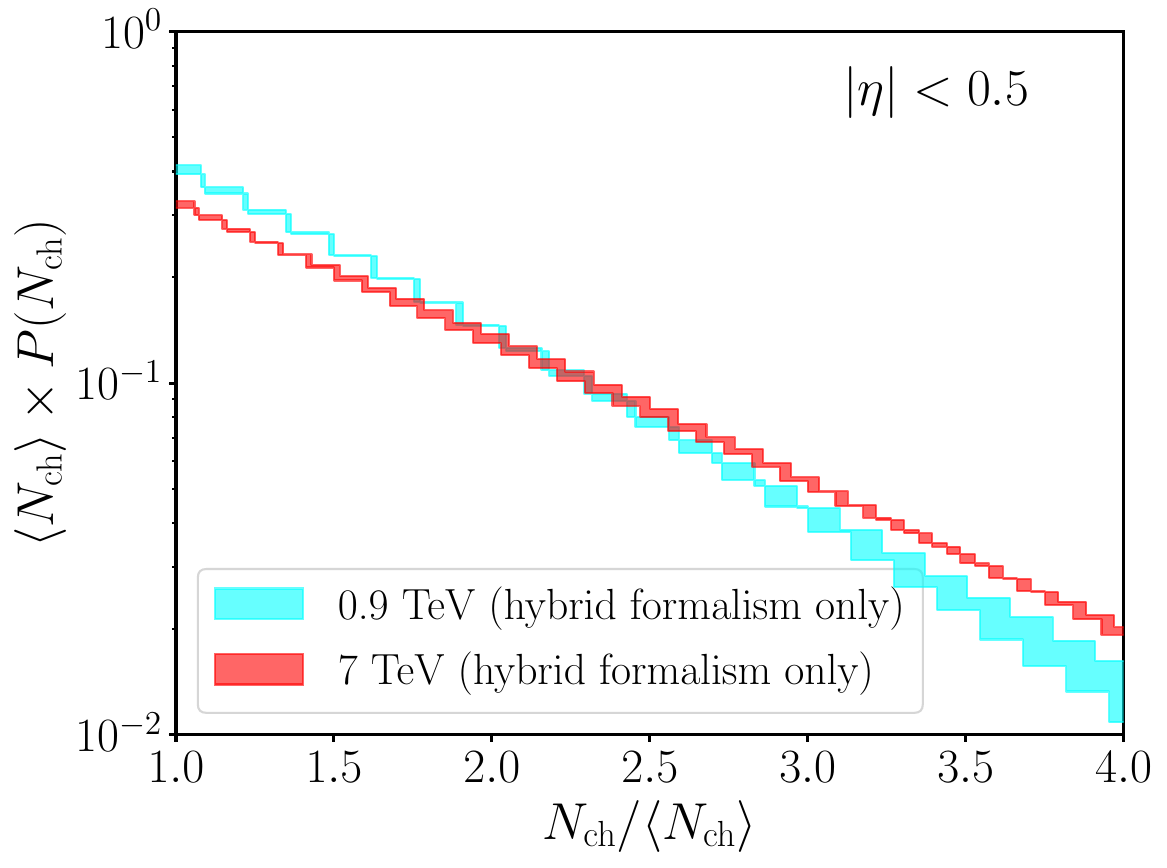}& 
\includegraphics[scale=0.4]{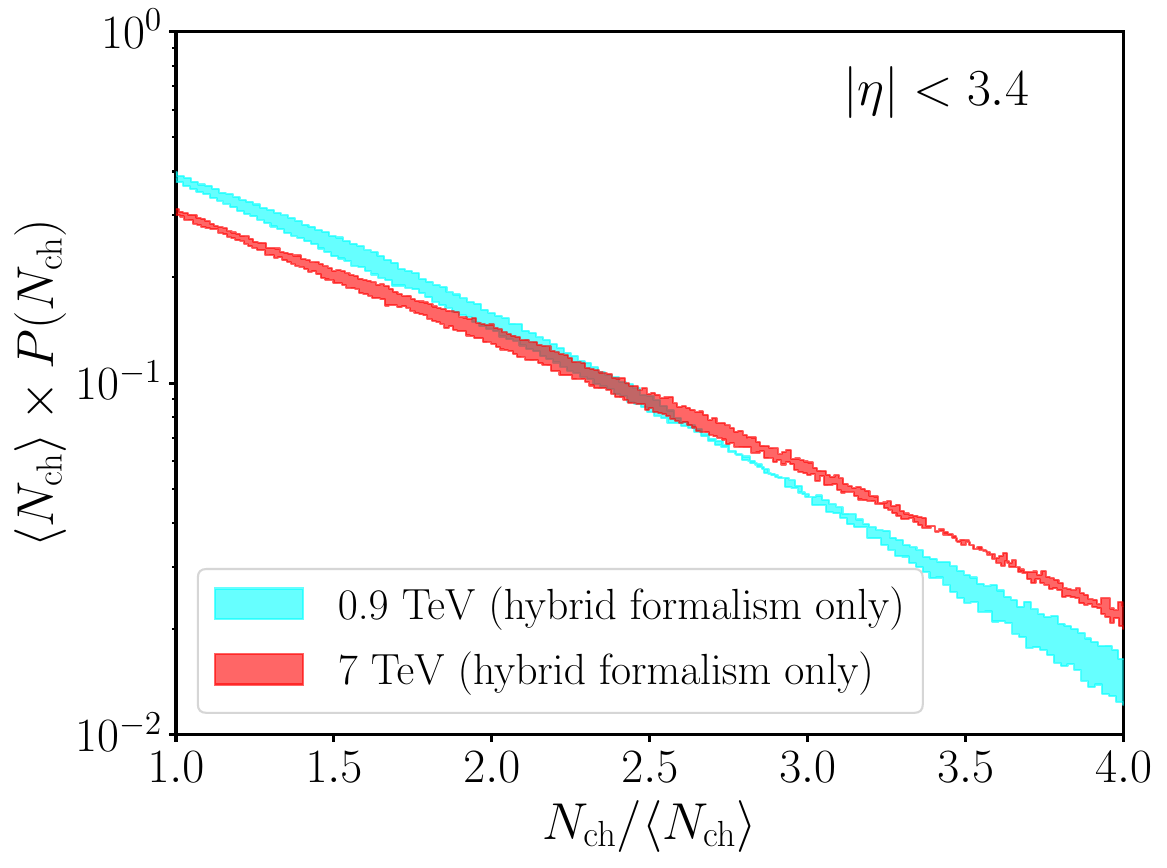} \\
(c) & (d)\\
\end{tabular}
\caption{KNO plot from the $k_T$-factorization formalism for a) $|\eta|<0.5$ and b) $|\eta|<3.4$. KNO plot from the hybrid formalism for c) $|\eta|<0.5$ and d) $|\eta|<3.4$.}
\label{kno-comp}
\end{figure}

In Figs. \ref{kno-comp}a and \ref{kno-comp}b we show the theoretical results 
obtained with the $k_T$-factorization  approach only, i.e., with the same   
parameters from Table \ref{parameters}, but without the weight function. It 
can be seen that for $|\eta|<0.5$ the results present approximate 
KNO scaling, but for a wider window ($|\eta|<3.4$), it is violated. 
In Figs. \ref{kno-comp}c and \ref{kno-comp}d  we 
show the results obtained  with hybrid formalism only. KNO scaling is 
never satisfied in this formalism, not even for $|\eta|<0.5$.

The results presented in those figures indicate that in the intermediate 
multiplicity region, KNO scaling occurs in the regime where small-x gluons 
dominate the dynamics, implemented in our work through the $k_T$-factorization 
formalism. As the pseudorapidity increases, the scaling is broken both by the  
$k_T$-factorization and hybrid formalism contributions simultaneously. 
In other words, KNO scaling is achieved by interactions dominated by small-x 
gluons in the central region, but violated by contributions of processes 
involving small-x gluons, collinear gluons and valence quarks in the 
fragmentation region.

 \section{Conclusions}

In this work we have combined two CGC models of particle production: the
$k_T$-factorization and the hybrid formalism. The former is appropriate for 
particle production in the central  pseudorapidity region 
whereas the latter is valid for particle production in large pseudorapidity
region (fragmentation region). These two formalisms were connected through a 
smooth function which gradually switches off $k_T$-factorization and switches 
on the hybrid formalism. As a result,  it was possible to 
reproduce the experimentally measured distributions over the whole 
pseudorapidity space. Then, with this two-component model we reproduced the 
measured multiplicity distributions in  pp collisions. In our model the dynamics 
in the central region is different from the fragmentation region and this 
difference manifested itself in the multiplicity distributions. In particular
we found KNO scaling in the central region and KNO violation in the 
fragmentation region. We were then able to identify the source of KNO 
scaling violation. This led us to the conjecture that collinear gluons and 
valence quarks from the fragmentation region may be responsible for KNO scaling 
violation.  Our analysis shows that this is true but the distributions obtained 
with the $k_T$-factorization (central region) formalism also 
show KNO violation for large rapidity windows. Our conclusion is then that the 
scaling is broken by simultaneous contributions of processes involving small-x gluons, 
collinear gluons and valence quarks.

\begin{acknowledgments}

We are grateful to Y. Lima for useful discussions. This study was partially 
financed  by the São Paulo Research Foundation       
(FAPESP), Brasil. Process Number 2024/06652-4. This work was also financed by the 
Brazilian funding agencies CNPq, FAPESP, CAPES, FAPERJ and by the INCT-FNA. 
This work has been done as a part of the Project INCT-Física
Nuclear e Aplicações, Projeto No. 408419/2024-5.

The authors acknowledge the National Laboratory for Scientific Computing  
(LNCC/MCTI, Brazil), through the ambassador program (UFGD), subprojects 
FCNAE and SADFT for providing HPC resources of the SDumont supercomputer. 

A.V. Giannini acknowledges support from CNPq through Universal Grant 405458/2025-8.

\end{acknowledgments}


\begin{thebibliography}{99}

\bibitem{mult1} W. Kittel, E.A. De Wolf,
Soft Multihadron Dynamics, World Scientific, (2005).

\bibitem{mult2} J. F. Grosse-Oetringhaus, K. Reygers,
J. Phys. G {\bf 37}, 083001 (2010).
arXiv:2506.09899.

\bibitem{KNO} Z. Koba, H.B. Nielsen and P. Olesen,
Nucl. Phys. B {\bf 40}, 317 (1972).


\bibitem{Braun:1998ty}
M.~A.~Braun and C.~Pajares,
Phys. Lett. B \textbf{444}, 435 (1998).

\bibitem{cms11} V.~Khachatryan [CMS],
JHEP \textbf{01}, 079 (2011); [arXiv:1011.5531 [hep-ex]].

\bibitem{ALICE_17} J. Adam \textit{et al.} [ALICE Collaboration],
Eur. Phys. J. C {\bf 77}, 33 (2017).

\bibitem{atlas22} Y.~Kulchitsky and P.~Tsiareshka,
Eur. Phys. J. C \textbf{82}, 462 (2022).

\bibitem{atlas23} Y.~A.~Kulchitsky and P.~Tsiareshka,
JHEP \textbf{10}, 111 (2023).

\bibitem{UA5} UA5 Collaboration,
Phys. Lett. B {\bf 160}, 193 (1985).

\bibitem{germano2} G.~R.~Germano and F.~S.~Navarra,
Phys. Rev. D \textbf{105}, 014005 (2022).

\bibitem{Mueller} A.H. Mueller,
Nucl. Phys. B {\bf 241}, 141 (1984).

\bibitem{Dremin} I.M. Dremin,
Sov. J. Part. Nucl. Phys. {\bf 22}, 37 (1991).

\bibitem{Dokshitzer} Yu.L. Dokshitzer, V.A. Khoze, A.H. Mueller
and S.I. Troyan, \textit{Basics of Perturbative QCD},
Editions Frontieres, Gif-sur-Yvette (1991).

\bibitem{gio86} A. Giovannini and L. Van Hove,
Z. Phys. C {\bf 30}, 391 (1986).

\bibitem{mpi} Y.~N.~Lima, C.~Jahnke, M.~G.~Munhoz and F.~S.~Navarra,
Phys. Rev. D \textbf{114}, 014058 (2026).

\bibitem{DPM} A. Capella, U. Sukhatme, C.I. Tan and J. Tran Thanh Van,
Phys. Rept. {\bf 236}, 225 (1994).

\bibitem{thecgc} F. Gelis, E. Iancu, J. Jalilian-Marian and R. Venugopalan,
Ann. Rev. Nucl. Part. Sci. {\bf 60}, 463 (2010).

\bibitem{saturation_physics} J. Jalilian-Marian and Y. Kovchegov,
Prog. Part. Nucl. Phys. {\bf 56}, 104 (2006).

\bibitem{dumi} A. Dumitru and E. Petreska, Phys. Rev.
D {\bf 83}, 114003 (2011).

\bibitem{vanhove} L.~Van Hove and S.~Pokorski,
Nucl. Phys. B \textbf{86}, 243 (1975).

\bibitem{fowler} G.~N.~Fowler, F.~S.~Navarra, M.~Plumer,
A.~Vourdas, R.~M.~Weiner and G.~Wilk,
Phys. Rev. C \textbf{40}, 1219 (1989).

\bibitem{ALICE_pp_eta_08} S. Acharya et al. [ALICE Collaboration], Phys. Lett. B {\bf 853}, 138700 (2024).

\bibitem{ALICE_wide} S. Acharya \textit{et al.} [ALICE Collaboration],
Eur. Phys. J. C {\bf 77}, 852 (2017).

\bibitem{henrique} H. R. Martins-Fontes and F. S. Navarra,
Phys. Rev. D \textbf{112}, 094045 (2025);
MDPI Physics \textbf{7}, 57 (2025).

\bibitem{alba14} J. L. Albacete and C. Marquet,
Prog. Part. Nucl. Phys. {\bf 76}, 1 (2014).

\bibitem{khar05} D. Kharzeev, E. Levin and M. Nardi,
Nucl. Phys. A {\bf 747}, 609 (2005).

\bibitem{rcbk} J. L. Albacete, A. Dumitru, H. Fujii and Y. Nara,
Nucl. Phys. A {\bf 897}, 1 (2013).

\bibitem{hibr} A. Dumitru, A. Hayashigaki and J. Jalilian-Marian,
Nucl. Phys. A {\bf 765}, 464 (2006).

\bibitem{pA_karzeev} A.Dumitru, D. Kharzeev, E. Levin, Y. Nara, Phys. Rev. C {\bf 85}, 044920 (2012).

\bibitem{dunga} F. Durães, A. V. Giannini, V. P. Gonçalves and F. S. Navarra, Phys. Rev. C {\bf 94}, 024917 (2016).

\bibitem{nara_26} H. Fujii, T. Hirano, K. Itakura, Y. Nara, S. Zhao, arXiv:2605.15494 [hep-ph].

\bibitem{yuri_D} Y. N. Lima, A. V. Giannini and V. P. Gonçalves, Phys. Rev. D {\bf 109}, 094035 (2024).

\bibitem{kov02} Y. V. Kovchegov and K. Tuchin,
Phys. Rev. D {\bf 65}, 074026 (2002).

\bibitem{Miller:2007ri}
M.~L.~Miller, K.~Reygers, S.~J.~Sanders and P.~Steinberg,
Ann. Rev. Nucl. Part. Sci. \textbf{57}, 205 (2007).



\bibitem{yuri_po} Y. Lima, L. J.F. Silva, A. V. Giannini and M. Munhoz
[arXiv:2602.11452v1 [hep-ph]].

\bibitem{ct14} S. Dulat \textit{et al.},
Phys. Rev. D {\bf 93}, 033006 (2016).

\bibitem{larry16} L.~McLerran and P.~Tribedy,
Nucl. Phys. A \textbf{945}, 216 (2016).

\bibitem{tgn26} R. Terra, A. V. Giannini and F. S. Navarra,
Phys. Rev. D {\bf 113}, 094009 (2026).

\bibitem{tgn-d} R. Terra, A. V. Giannini and F. S. Navarra
[arXiv: 2603.24344 [hep-ph]].

\bibitem{ipjazzma} J. L. Nagle and W. A. Zajc,
Phys. Rev. C {\bf 99}, 054908 (2019).

\bibitem{schenke} H. Mantysaari and B. Schenke,
Phys. Rev. D \textbf{94}, 034042 (2016).

\bibitem{running_the_gamut} B. Schenke, C. Shen and P. Tribedy, Phys. Rev. C {\bf 102}, 044905 (2020).

\bibitem{bp_raju} A. H. Rezaeian, M. Siddikov, M. Van de Klundert and R. Venugopalan, Phys. Rev. D {\bf 87}, 034002 (2013).


\bibitem{ALICE_5} S. Acharya \textit{et al.} [ALICE Collaboration],
Phys. Rev. D {\bf 108}, 072008 (2023).

\bibitem{TOTEM7} G. Antchev \textit{et al.} [TOTEM Collaboration],
EPL {\bf 98}, 31002 (2012).

\bibitem{TOTEM8} S. Chatrchyan \textit{et al.} [CMS and TOTEM
Collaborations], Eur. Phys. J. C {\bf 74}, 3053 (2014).

\bibitem{mean_pt} V. Khachatryan  \textit{et al.} [CMS Collaboration], Phys. Rev. Lett. {\bf 105}, 022002  (2010).


\bibitem{LHCb_pt} R. Aaij \textit{et al.} [LHCb Collaboration],
 [arXiv:2607.26783v3 [nucl-ex]] .

\end{thebibliography}
\end{document}